\documentclass{aa}
\usepackage[varg]{txfonts}

\usepackage{graphicx}

\usepackage{natbib}

\def\apjl{ApJL }

\def\apjs{ApJS }
\def\aap{A\&A }

\bibpunct{(}{)}{;}{a}{}{,} 

\begin{document} 

\titlerunning{VLT/UVES observations of peculiar $\alpha$ abundances in a sub-DLA at $z\approx 1.8$}
\title{VLT/UVES observations of peculiar $\alpha$ abundances in a sub-DLA at $z\approx 1.8$ towards 
the quasar B1101$-$26}

\authorrunning{A. Fox, P. Richter, C. Fechner}

\author{Anne Fox\inst{\ref{inst1}}, Philipp Richter\inst{\ref{inst1}}, \and Cora Fechner\inst{\ref{inst1}}}

\institute{Institut für Physik und Astronomie, Karl-Liebknecht-Str. 24/25, 14476 Potsdam, Germany}\label{inst1} 

\date{Draft version September 29, 2014}

\abstract {We present a detailed analysis of chemical abundances in a sub-damped Lyman $\alpha$ absorber at 
$z=1.839$ towards the quasar B1101$-$26, based on a very-high-resolution ($R\sim 75,000$) and high-signal-to-noise 
(S/N $>100$) spectrum observed with the UV Visual Echelle spectrograph (UVES) installed on the ESO Very Large Telescope (VLT). 
The absorption line profiles are resolved into a maximum of eleven velocity components spanning a rest-frame velocity 
range of $\approx 200$\,km s$^{-1}$. Detected ions include C\,{\sc ii}, C\,{\sc iv}, N\,{\sc ii}, O\,{\sc i}, Mg\,{\sc i},
Mg\,{\sc ii},  Al\,{\sc ii}, Al\,{\sc iii}, Si\,{\sc ii}, Si\,{\sc iii}, Si\,{\sc iv}, Fe\,{\sc ii}, 
and possibly S\,{\sc ii}. The total neutral hydrogen column density is $\log N($H\,{\sc i}$) = 19.48 \pm 0.01$. 
From measurements of column densities and Doppler parameters we estimate element abundances of the above-given
elements. The overall metallicity, as traced by [O\,{\sc i}/H\,{\sc i}], is $-1.56 \pm 0.01$. For the nitrogen-to-oxygen ratio 
we derive an upper limit of [N\,{\sc i}/O\,{\sc i}] $\le -0.65$, which suggests a chemically young absorption line system. This is supported 
by a supersolar $\alpha$/Fe ratio of [Si\,{\sc ii}/Fe\,{\sc ii}] $\approx 0.5$. 
The most striking feature in the observed abundance pattern is an unusually high 
sulphur-to-oxygen ratio of $0.69 \le $ [S\,{\sc ii}/O\,{\sc i}] $\le 1.26$. We calculate detailed photoionisation models for two subcomponents with Cloudy, and can rule out that ionisation effects alone are responsible for the high S/O ratio. We instead speculate that
the high S/O ratio is caused by the combination of several effects, such as specific ionisation 
conditions in multi-phase gas, unusual relative abundances of heavy elements, and/or dust depletion in a local
gas environment that is not well mixed and/or that might be related to star-formation activity in the host galaxy. We discuss the implications of our findings for the interpretation of $\alpha$-element abundances in metal absorbers at high redshift.}


\maketitle

\section{Introduction}

The formation and evolution of galaxies are important aspects for our understanding of the
Universe and its structure. In particular, the reconstruction of the chemical history of galaxies and the 
metal content in the early Universe are fundamental steps to understanding these basic processes. 
One efficient way to investigate them is to study the distribution and chemical composition of gas inside
and outside of galaxies in the interstellar and intergalactic medium (ISM and IGM) at different redshifts. This gas can be studied through intervening absorption line systems detected in the spectra of bright background sources, for example quasars. This approach provides the advantage of being unbiased with respect to distance, luminosity, and morphology of the host galaxies, whereas the observation of the faint starlight is exceedingly difficult. 

Intervening absorbers are classified into four categories according to their neutral hydrogen column density (and, consequently, their relation to galaxies). Intergalactic gas clouds with column densities of neutral hydrogen less than $N$(H\,{\sc i}) $\lesssim 10^{17}$\,cm$^{-2}$ are called Lyman $\alpha$ forest systems \citep[see e.g.][]{Meiksin2009}. These absorbers are believed to predominantly trace the low-density intergalactic medium (IGM) beyond the virial radii of galaxies. The so-called Lyman limit systems (LLS) have column densities up to $10^{19}\mathrm{\,cm}^{-2}$ and are believed to
mostly arise in the the circumgalactic medium (CGM) of galaxies. These systems might be important tracers
of gas infall and outflow processes that circulate the gas around galaxies \citep{Crighton2013,vandeVoort2012,Rudie2012,Faucher-Giguere2011,Fumagalli2011} and could therefore be a significant metal reservoir \citep{Prochaska2006}. 
Absorption line systems with neutral hydrogen column densities $>2\times 10^{20}$ cm$^{-2}$ are called damped Lyman $\alpha$ systems (DLAs). 
The sub-damped Lyman $\alpha$ systems (sub-DLAs) comprise the transition between Lyman limit systems and DLAs with column densities $10^{19}$ 
cm$^{-2} \le N($H\,{\sc i}$) \le 2 \times 10^{20}$ cm$^{-2}$ \citep{Dessauges2003,Peroux2003}. Because of their large 
neutral hydrogen content, DLAs and sub-DLAs are believed to be associated with the inner regions of galaxies (e.g. with
the gas disks of spirals). 

Absorption-line measurements indicate that DLA systems dominate the neutral gas content of the Universe 
at $z>1$ \citep{Lanzetta1995,Rao2000}. Studies of DLAs  are in line with the idea that these systems are the progenitors of present-day galaxies \citep[e.g.][]{Moller2013,Rafelski2012,Rafelski2011,Yin2011,Zwaan2005,Wolfe2005,Boissier2003,Alvensleben2001,Salucci1999,Prochaska1997}. 
In contrast to DLAs, sub-DLAs have lower neutral gas column densities, but they are generally considered to be more metal rich than DLAs \citep{Som2013,Meiring2009,Peroux2007}. However, ionisation corrections may be significant for these systems and notably alter the metallicity estimates. Nonetheless, for a given metallicity, sub-DLAs have lower metal column densities than DLAs owing to the low H\,{\sc i} column densities. This allows us to accurately measure the abundance of the most common elements like oxygen and carbon without facing the problem of line saturation. Accurate abundance measurements of metals are particularly important to learn about the chemical enrichment of the gas inside and outside of galaxies, because they constrain the conditions for early nucleosynthesis in the Universe.

As certain types of enrichment processes cause a specific abundance pattern in the ISM and IGM, we can deduce different types 
of stellar explosions and physical processes from the abundance measurements in absorption line systems. In order to explain the metal-abundance 
patterns in DLAs, two complementary effects are often brought up: (1) the nucleosynthetic enrichment from 
Type II supernovae (SNe) and (2) an interstellar medium-like (ISM-like) dust-depletion pattern. A typical Type II SN 
enrichment pattern shows an overabundance of O relative to N, an underabundance of elements with odd atomic number compared to elements with even atomic number (e.g. Mn/Fe and Al/Si), and an overabundance of $\alpha$-elements (e.g. O, Si, S) relative to iron peak elements (e.g. Cr, Mn, Fe, Ni) \cite[see][and references therein]{Lu1996}. The last is caused by the relatively quick production of $\alpha$-capture elements and some iron peak elements in Type II SNe, while at later times a dominant production of the iron peak elements comes from Type Ia SNe whose precursors evolve more slowly than the massive stars that cause SNe Type II. The production of nitrogen, on the other hand, is less well understood, but it is assumed that the production is dominated by intermediate-mass stars \citep{Henry2000}.

Comprehensive studies of DLAs, performed for example by \citet{Lu1996} and \citet{Prochaska1999}, found abundance patterns 
that are broadly consistent with a combination of SN Type II enrichment and dust-depletion, but at the same time they find 
exceptions that cannot be explained by these two effects. \citet{Som2013} confirmed suggestions for sub-DLAs being more 
metal-rich than DLAs and evolving faster. Besides, abundance ratios of [Mn/Fe] in their study suggest a difference in 
the stellar populations for the galaxies traced by DLAs and sub-DLAs. It seems that, even though many observational studies of 
DLAs and sub-DLAs have been pursued \citep[e.g.][]{Ledoux2003,Dessauges2003,Som2013}, the abundance patterns lack a 
convincing single interpretation. Furthermore, the characteristic of the influence of dust-depletion on the observed abundance 
ratios \citep{Vladilo2011} is an ongoing issue of debate.

Since the origin of sub-DLAs with respect to the location of the star-forming inner regions of galaxies is unclear, 
these systems have been largely ignored until recently. However, detailed studies of this class of absorbers are advantageous and 
desirable, not only for the observational reasons mentioned above. One major drawback that hampers the interpretation
of chemical abundances in LLS, sub-DLAs and DLAs is the often complex velocity-component structure in the absorbers.
This velocity structure reflects the complex spatial distribution of the different gas phases in and around galaxies
at pc and kpc scales. Since a typical quasar sightline that contains a sub-DLA or DLA passes both the inner (star-forming)
region of a galaxy and the outer circumgalactic environment, it cannot be readily expected that the different absorption
components contain gas with one and the same chemical enrichment pattern. While most of the previous studies
of DLAs and sub-DLAs largely ignore this aspect, partly because of limitations in spectral resolution and
signal-to-noise (S/N) in the data, some detailed studies on kinematically complex DLAs and sub-DLAs at high $z$ 
indeed indicate a non-uniform abundance pattern within the absorbers \citep{Crighton2013,Prochter2010,D'Odorico2001}, e.g. as part of merger
processes \citep{Richter2005}. Such accurate measurements require very high-resolution spectra of quasars with high S/N ($ \gtrsim 50$), but these are sparse and detailed analyses of their absorption line systems are even rarer. 

To improve our understanding of the chemical enrichment pattern within DLAs and sub-DLAs at high redshift, we present in this 
paper a detailed analysis of a complex sub-DLA at $z=1.839$ in the optical spectrum of the quasar
B1101$-$26. Our analysis is based on a very-high-spectral-resolution ($R=75\,000$) and high-signal-to-noise ratio (S/N $\sim 100$) 
spectrum of QSO B1101$-$26 taken with VLT/UVES. After presenting our data set and analysis method in 
Sec.\ \ref{Observations, data handling and analysis method}, we present the results of the absorption-line fitting and the photoionisation models 
in Sec.\ \ref{Results and discussion}.
In Sec.\ \ref{Discussion}, we give a detailed discussion of our findings and summarise our study in Sec.\ \ref{Summary}. 
Throughout the paper we use the list of atomic data provided by \citet{Morton2003}. 

\section{The sub-DLA towards QSO B1101$-$26}

The optical spectrum of QSO B1101$-$26, (other name: 2MASS\,J110325$-$264515, $m_{V}=16\fm02$, $z_{\text{em}}=2.145$) 
contains a prominent sub-DLA at $z_{\text{abs}}=1.839$ that exhibits a complex velocity-component structure
spanning more than $200$ km\,s$^{-1}$ in the absorber restframe.
This sub-DLA has been subject of earlier studies, e.g. by \citet{Petitjean2000} and by \citet{Dessauges2003}. 
These previous studies used early observational data at high spectral resolution ($R\sim 45,000$) 
obtained in 2000 with the UV-Visual Echelle Spectrograph (UVES) installed on ESO Very Large Telescope (VLT) as part of the UVES Science Verification. 
From the analysis of these data, \citet{Dessauges2003} derived a total neutral hydrogen column density of 
$\log N$(H\,{\sc i}) $=19.50 \pm 0.05$ in the sub-DLA from a fit of the Ly\,$\alpha$ line. In their dataset, that has an
average velocity resolution of $5-7$ km\,s$^{-1}$, they could resolve the absorption pattern of the weakly ionised species 
into 11 subcomponents, whereas the highly ionised species have been resolved into six subcomponents. 
Moreover, \citet{Dessauges2003} have measured the column densities in the sub-DLA for 11 ions and seven elements (Si, O, S, C, Al, Fe, Mg) 
that they could identify. For all these elements, they have computed the abundances using solar abundances provided by \citet{Grevesse1998}.
For example, they have confirmed the measurements of \citet{Petitjean2000} for the iron abundance being [Fe/H]$=-1.49\pm$0.05 and they 
have determined an oxygen (= alpha) abundance of [O/H]$=-1.78\pm0.12$ that surprisingly is significantly {\it lower} than that of Fe. However, the true velocity structure in the gas is possibly not fully resolved at the given spectral resolution. Concludingly, the true Doppler parameters ($b$-values) in the individual (unresolved) components may be significantly lower than what has been estimated previously and, thus, the column densities could be underestimated.
Therefore, the limited spectral resolution and S/N leads to systematic uncertainties for the derived abundances. A more detailed analysis of this complex absorption line system with higher-resolution data is possible with the data set used here and can provide deeper insights into the metal enrichment at high redshift, based on more accurate abundance measurements.

\section{Observations, data handling, and analysis method}\label{Observations, data handling and analysis method}

\begin{figure*}
 \centering
\includegraphics[width=17cm,viewport = 80 470 540 790,clip]{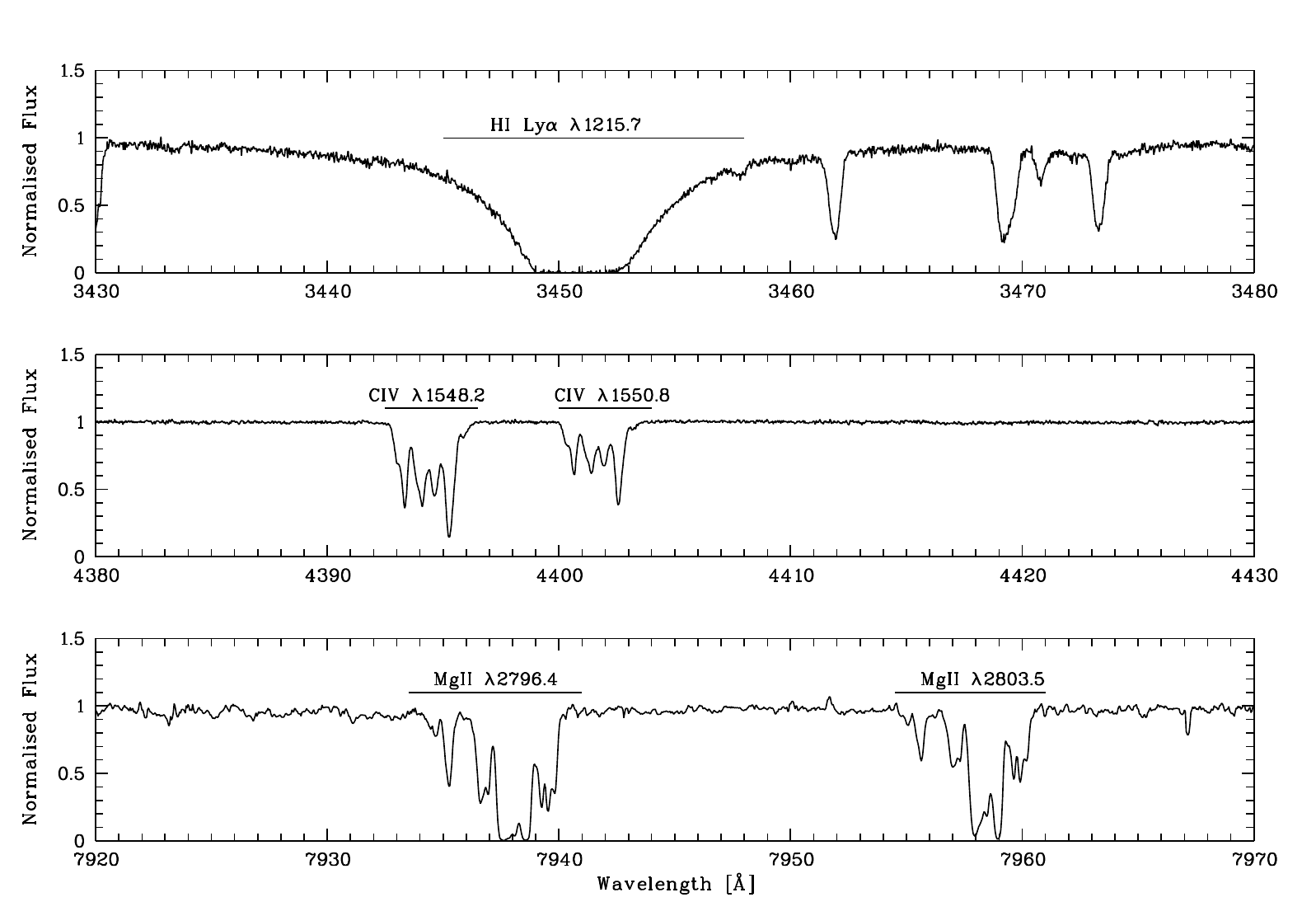} 
\caption{Portion of UVES spectrum of QSO B1101$-$26. In the upper panel, the H\,{\sc i} 
Ly\,$\alpha$ absorption in the sub-DLA at $z=1.839$ is shown. The middle and lower panel 
show the C\,{\sc iv} and Mg\,{\sc ii} absorption lines of the same absorption line system, respectively.}
\label{spektrumauszug}
\end{figure*}

The spectral data used in this study have been obtained with UVES/VLT in February 2006 as part of the VLT observing programme 076.A-0463(A). 
This programme aims at constraining the cosmological variability of the fine-structure constant by observing metal absorption-line systems 
in quasar spectra (PI/Col: Lopez/ Molaro/ Centuri\'on/ Levshakov/ Bonifacio/ D'Odorico). QSO B1101$-$26 was observed through a $0.5$ arcsec slit 
under good seeing conditions ($\lesssim 1$ arcsec) during the run. The total integration time was 100\,261\,s. The data are of high spectral resolution 
($R\sim75\,000$ or 4 km s$^{-1}$ FWHM) and exhibit an excellent S/N of $\sim100$, on average ($45$ at $3550$\,\AA, $140$ at $4400$\,\AA). 
The raw data were reduced and normalised to the local continuum of the quasar as part of the UVES Spectral Quasar Absorption Database 
(SQUAD; PI: Michael T. Murphy) using a modified version of the UVES pipeline. The reduction includes flat-fielding, 
bias- and sky-subtraction, and a relative wavelength calibration. The final combined spectrum covers a wavelength range of $3050-10\,430$\,\AA. 
A representative portion of the UVES spectrum of QSO B1101$-$26 is displayed in Fig.\ \ref{spektrumauszug}, showing selected absorption 
lines of the sub-DLA system at $z=1.839$. The data are publicly available in the UVES database within ESO's Science Archive 
Facility\footnote[1]{\texttt{http://archive.eso.org}}.

The high resolution and high S/N of the new UVES spectrum of QSO B1101$-$26 enables us to identify absorption lines arising 
from 14 different ions, namely H\,{\sc i}, C\,{\sc ii}, C\,{\sc iv}, N\,{\sc ii}, O\,{\sc i}, Mg\,{\sc i}, Mg\,{\sc ii}, Al\,{\sc ii}, Al\,{\sc iii}, Si\,{\sc ii}, Si\,{\sc iii}, Si\,{\sc iv}, Fe\,{\sc ii}, and possibly S\,{\sc ii}. 
We have analysed the spectrum using the \texttt{FITLYMAN} package implemented in MIDAS \citep{Fontana1995}. 
This routine uses a $\chi^{2}$-minimisation algorithm for multi-component Voigt-profile fitting to derive 
column densities, $N$, and Doppler-parameters, $b$, taking into account the instrumental line-spread function (LSF). Because different lines from the same ion are fitted simultaneously, $b$-values obtained from the fit can be smaller than the velocity resolution of the instrument, as the information on the relative line strengths enters the determination of $b$ in the fit. This also means that even at this high spectral resolution the absorption profiles may be unresolved and the true velocity-component structure remains uncertain. Consequently, the physical interpretation of the derived $b$-values (e.g. in terms of thermal 
broadening processes) remains afflicted with systematic uncertainties.

\section{Results and discussion}\label{Results and discussion}

\subsection{Velocity structure}

\begin{figure*}
 \centering
\includegraphics[width=15cm,viewport = 5 80 540 790,clip]{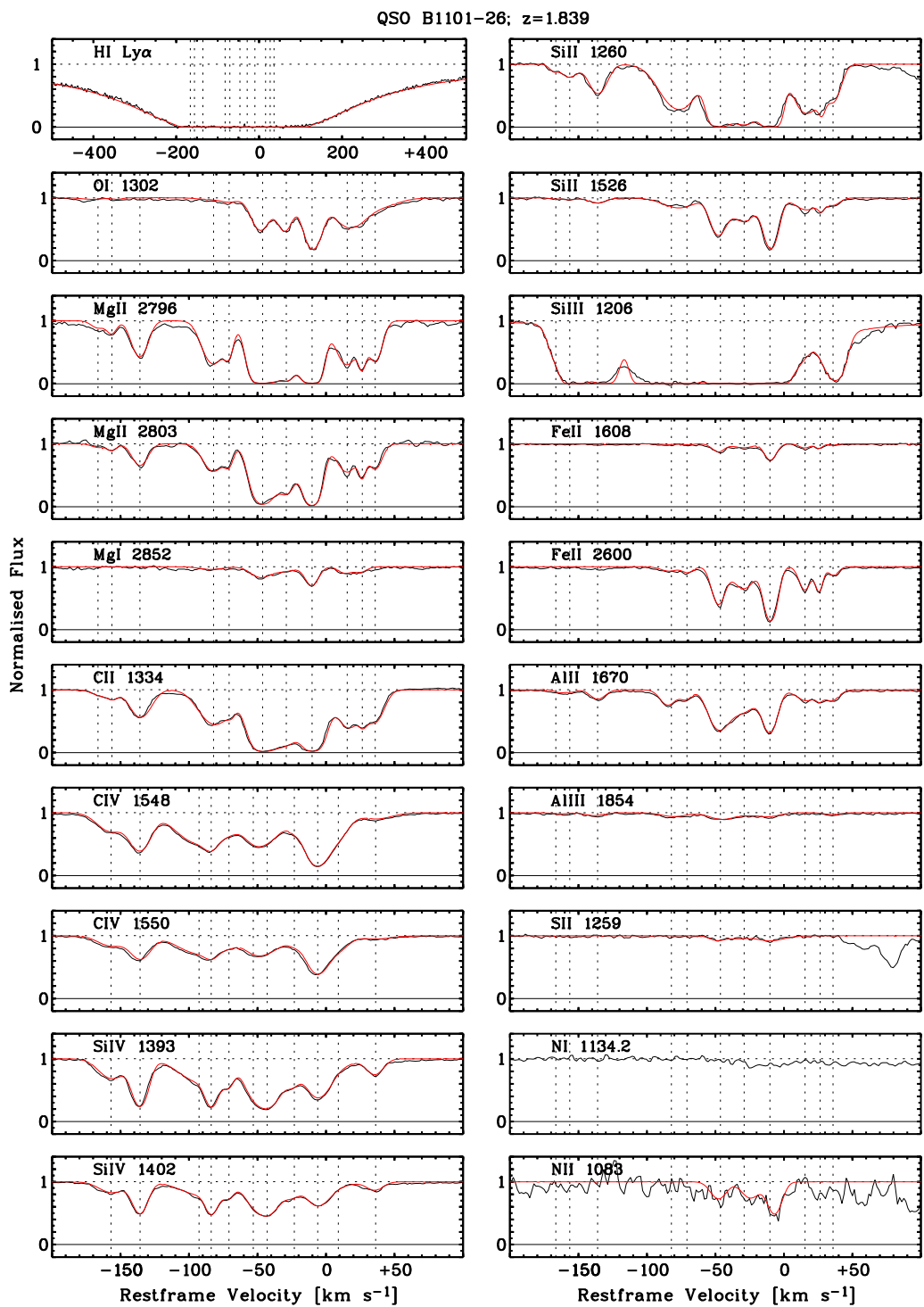} 
\caption{Absorption profiles of H\,{\sc i}, C\,{\sc ii}, C\,{\sc iv}, N\,{\sc i}, N\,{\sc ii}, O\,{\sc i}, Mg\,{\sc i}, Mg\,{\sc ii}, Al\,{\sc ii}, Al\,{\sc iii}, Si\,{\sc ii}, Si\,{\sc iv},  S\,{\sc ii},  and  Fe\,{\sc ii}  for the $z=1.839$ sub-DLA. The individual absorption components are indicated by the vertical dashed lines. The velocity scale refers to the $z=1.839$ restframe. The multi-component Voigt profile fits are overlayed in red. Parameters of the fit can be found in Table\ \ref{Fit results}.}
\label{fitovervelocityscale}
\end{figure*}

\begin{figure}
 \resizebox{\hsize}{!}{\includegraphics{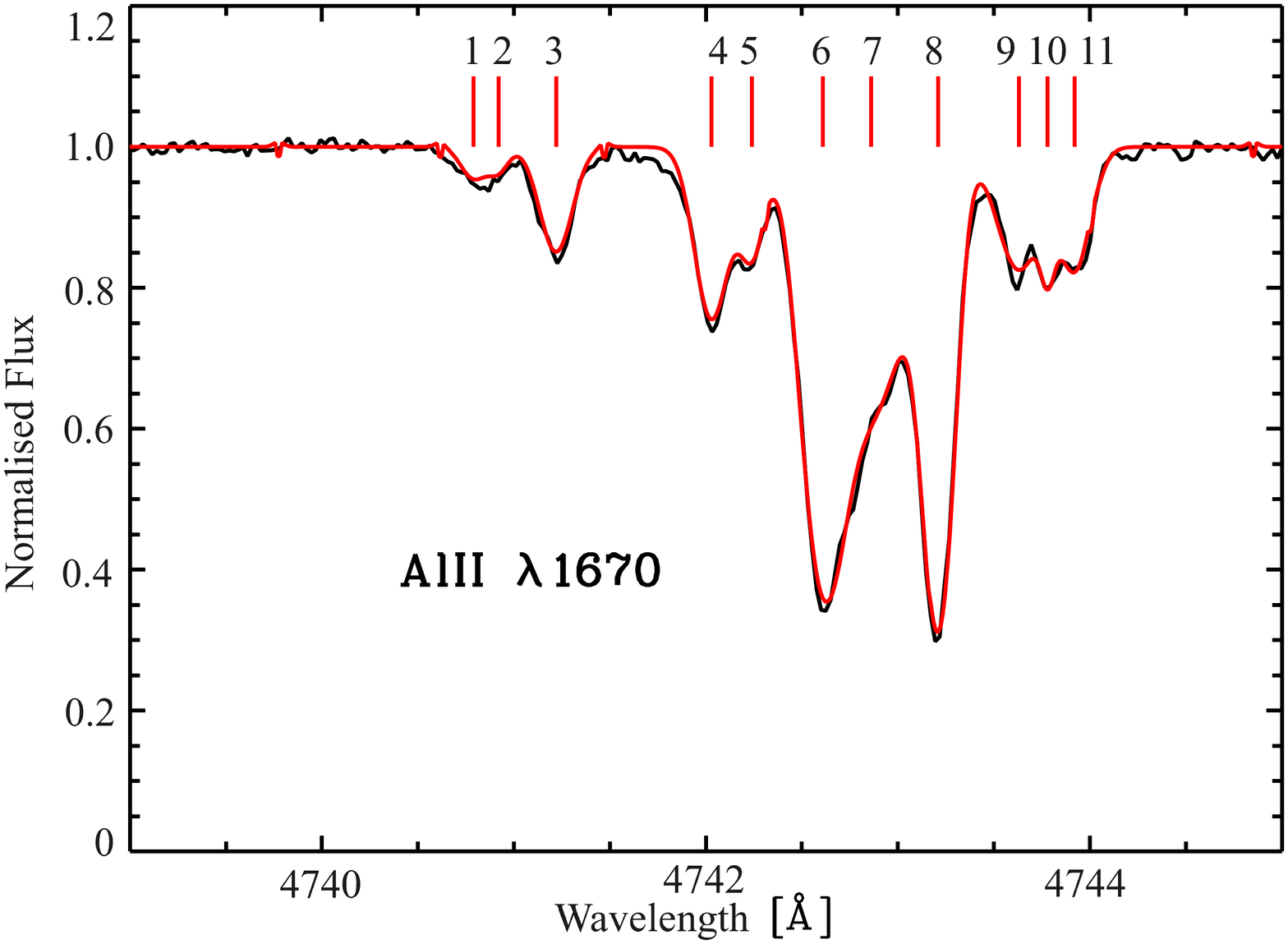}}
\caption {Representative multi-component Voigt profile fit of the Al\,{\sc ii} $\lambda 1670.79$\,\AA\ absorption feature  in the sub-DLA at $z=1.839$. 
The data and the fit are indicated in black and red. The velocity components 
used in the fits are indicated by vertical tick marks. Parameters of the fit can be found in 
Table\ \ref{Fit results}.} \label{FitAlII1670}
\end{figure}

The component structures in the $z=1.839$ absorber for neutral and weakly ionised species and highly ionised species 
were analysed separately, because these ions usually do not arise in the same gas phase \citep{Crighton2013,Milutinovic2010,Fox2007b,Wolfe2005} (see Fig.\ \ref{fitovervelocityscale}). The Mg\,{\sc ii} absorption line is a good representative for the velocity structure of the weakly ionised species, 
because it represents a strong atomic transition that is found in a regime with a very high S/N and is mostly unaffected 
by blending. Therefore, we have used the velocity-component structure of Mg\,{\sc ii} as template for all weakly ionised species 
to obtain comparable column densities for the individual components, but we allow a $\Delta \lambda = 0.05$\,\AA\ wavelength displacement ($\simeq 2$ pixels) in the absolute positions of the absorption components during the fitting process. Because of the large number of absorption lines and components, a simultaneous fit for all of them has not been possible. However,  multiplets of the same ion have been fitted simultaneously, which allows us to determine the fitting 
parameters with high accuracy. For weak or undetected absorption features, we determine the limiting equivalent width 
based on the local S/N and considering a $3 \sigma$ detection limit for unsaturated line absorption. A representative 
multi-component fit of the Al\,{\sc ii} $\lambda$1670 absorption line is shown in Fig.\ \ref{FitAlII1670}. 
The highly ionised species C\,{\sc iv} and Si\,{\sc iv} have been fitted simultaneously altogether. Therefore, they 
have a common velocity structure by definition. 

\clearpage

\begin{table}[h]
\centering
\scriptsize
\caption{Component structure, column densities, and Doppler parameters for the $z=1.839$ absorber towards QSO B1101$-$26 \label{Fit results}}
\begin{tabular}{ccccc} \hline\hline
\noalign{\vspace{1mm}}
No&$v_{\text{rel}}$ [km s$^{-1}$]& Ion A & $\log N$(A) & $b$ [km s$^{-1}$]\\
\noalign{\vspace{1mm}}
\hline
\noalign{\vspace{1mm}}
1&$-166.4$	&	C\,{\sc ii}&	 $12.61\pm0.05$	&	$9.5\pm0.6$	\\ 
&	&	N\,{\sc ii}&	 $-$	&	$-$	\\ 
&	&	O\,{\sc i}&	 $12.45\pm0.08(0.09)$	&	$10.7\pm2.3$	\\
&	&	Mg\,{\sc i}&	 $-$	&	$-$	\\ 
&	&	Mg\,{\sc ii}&	 $11.50\pm0.13$	&	$6.3\pm2.0$	\\ 
&	&	Al\,{\sc ii}&	 $10.79\pm0.06$	&	$5.3\pm0.9$	\\ 
&	&	Al\,{\sc iii}&	 $10.94\pm0.22(0.22)$	&	$7.9\pm4.5$	\\ 
&	&	Si\,{\sc ii}&	 $11.28\pm0.12$	&	$3.6\pm1.1$	\\ 
&	&	S\,{\sc ii}&	 $-$	&	$-$	\\ 
&	&	Fe\,{\sc ii}&	 $-$	&	$-$	\\ \hline
\noalign{\vspace{1mm}}

2&$-156.3$	&	C\,{\sc ii}&	 $12.35\pm0.09$	&	$4.8\pm0.5$	\\ 
&	&	N\,{\sc ii}&	 $-$	&	$-$	\\ 
&	&	O\,{\sc i}&	 $12.08\pm0.12(0.12)$	&	$4.2\pm1.1$	\\
&	&	Mg\,{\sc i}&	 $-$	&	$-$	\\
&	&	Mg\,{\sc ii}&	 $11.63\pm0.09$	&	$4.4\pm0.7$	\\ 
&	&	Al\,{\sc ii}&	 $10.55\pm0.09$	&	$3.9\pm1.0$	\\ 
&	&	Al\,{\sc iii}&	 $-$	&	$-$	\\ 
&	&	Si\,{\sc ii}&	 $11.94\pm0.04$	&	$7.9\pm0.9$	\\ 
&	&	S\,{\sc ii}&	 $-$	&	$-$	\\ 
&	&	Fe\,{\sc ii}&	 $-$	&	$-$	\\ \hline
\noalign{\vspace{1mm}}

3&$-135.9$	&	C\,{\sc ii}&	 $13.36\pm0.00$	&	$9.8\pm0.1$	\\ 
&	&	N\,{\sc ii}&	 $-$	&	$-$	\\ 
&	&	O\,{\sc i}&	 $-$	&	$-$	\\ 
&	&	Mg\,{\sc i}&	 $-$	&	$-$	\\ 
&	&	Mg\,{\sc ii}&	 $12.40\pm0.01$	&	$7.2\pm0.1$	\\ 
&	&	Al\,{\sc ii}&	 $11.44\pm0.01$	&	$7.0\pm0.2$	\\ 
&	&	Al\,{\sc iii}&	 $11.43\pm0.07(0.08)$	&	$8.9\pm1.7$	\\ 
&	&	Si\,{\sc ii}&	 $12.39\pm0.01$	&	$8.2\pm0.2$	\\ 
&	&	S\,{\sc ii}&	 $-$	&	$-$	\\ 
&	&	Fe\,{\sc ii}&	 $-$	&	$-$	\\ \hline
\noalign{\vspace{1mm}}

4&$-82.1$	&	C\,{\sc ii}&	 $13.67\pm0.01$	&	$14.6\pm0.2$	\\
&	&	N\,{\sc ii}&	 $-$	&	$-$	\\
&	&	O\,{\sc i}&	 $12.50\pm0.08(0.09)$	&	$11.0*$	\\ 
&	&	Mg\,{\sc i}&	 $10.80\pm0.06(0.13)$	&	$11.0*$	\\ 
&	&	Mg\,{\sc ii}&	 $12.65\pm0.02$	&	$9.6\pm0.4$	\\ 
&	&	Al\,{\sc ii}&	 $11.69\pm0.01$	&	$7.2\pm0.3$	\\ 
&	&	Al\,{\sc iii}&	 $11.69\pm0.05(0.06)$	&	$12.0\pm1.7$	\\
&	&	Si\,{\sc ii}&	 $12.37\pm0.13$	&	$17.4\pm0.8$	\\ 
 &	&	S\,{\sc ii}&	 $-$	&	$-$	\\ 
&	&	Fe\,{\sc ii}&	 $11.80\pm0.04(0.05)$	&	$7.9\pm1.1$	\\ \hline
\noalign{\vspace{1mm}}

5&$-70.8$	&	C\,{\sc ii}&	 $12.38\pm0.05$	&	$2.7\pm0.4$	\\
&	&	N\,{\sc ii}&	 $-$	&	$-$	\\ 
&	&	O\,{\sc i}&	 $12.81\pm0.05(0.06)$	&	$8.5\pm1.3$	\\
&	&	Mg\,{\sc i}&	 $10.20\pm0.17(0.21)$	&	$5.0*$	\\ 
&	&	Mg\,{\sc ii}&	 $12.06\pm0.05$	&	$3.7\pm0.3$	\\ 
&	&	Al\,{\sc ii}&	 $11.35\pm0.03$	&	$5.4\pm0.3$	\\ 
&	&	Al\,{\sc iii}&	 $-$	&	$-$	\\ 
&	&	Si\,{\sc ii}&	 $12.83\pm0.04$	&	$14.5\pm0.2$	\\
&	&	S\,{\sc ii}&	 $-$	&	$-$	\\ 
&	&	Fe\,{\sc ii}&	 $11.75\pm0.04(0.05)$	&	$5.0*$	\\ \hline
\noalign{\vspace{1mm}}

6&$-46.3$	&	C\,{\sc ii}&	 $14.22\pm0.03$	&	$10.3\pm0.3$	\\ 
&	&	N\,{\sc ii}&	 $13.14\pm0.07(0.07)$	&	$7.4*$	\\
&	&	O\,{\sc i}&	 $13.83\pm0.01(0.03)$	&	$8.3\pm0.2$	\\ 
&	&	Mg\,{\sc i}&	 $11.36\pm0.03(0.12)$	&	$9.6\pm0.7$	\\ 
&	&	Mg\,{\sc ii}&	 $13.37\pm0.02$	&	$8.8\pm0.2$	\\ 
&	&	Al\,{\sc ii}&	 $12.25\pm0.01$	&	$8.7\pm0.1$	\\ 
&	&	Al\,{\sc iii}&	 $11.96\pm0.05(0.06)$	&	$12.6\pm1.7$	\\
&	&	Si\,{\sc ii}&	 $13.33\pm0.01$	&	$7.0\pm0.1$	\\ 
&	&	S\,{\sc ii}&	 $13.28\pm0.08$	&	$7.4\pm1.5$	\\ 
&	&	Fe\,{\sc ii}&	 $12.78\pm0.01(0.03)$	&	$5.7\pm0.1$	\\ \hline
\noalign{\vspace{1mm}}
7&$-29.0$	&	C\,{\sc ii}&	 $13.88\pm0.06$	&	$9.7\pm1.1$	\\
&	&	N\,{\sc ii}&	 $13.17\pm0.07(0.07)$	&	$8.5*$	\\
&	&	O\,{\sc i}&	 $13.77\pm0.01(0.03)$	&	$6.9\pm0.2$	\\ 
&	&	Mg\,{\sc i}&	 $11.03\pm0.11(0.16)$	&	$10.9\pm2.8$	\\ 
&	&	Mg\,{\sc ii}&	 $13.01\pm0.03$	&	$8.0\pm0.6$	\\ 
&	&	Al\,{\sc ii}&	 $12.23\pm0.02$	&	$15.6\pm0.9$	\\ 
&	&	Al\,{\sc iii}&	 $11.13\pm0.47(0.47)$	&	$8.5*$	\\
&	&	Si\,{\sc ii}&	 $13.26\pm0.01$	&	$11.4\pm0.4$	\\
&	&	S\,{\sc ii}&	 $13.13\pm0.19$	&	$9.2\pm2.5$	\\ 
&	&	Fe\,{\sc ii}&	 $12.67\pm0.01(0.03)$	&	$9.8\pm0.3$	\\ \hline
\noalign{\vspace{1mm}}
\end{tabular}
\end{table}

\begin{table}
\centering
\scriptsize
\begin{tabular}{ccccc} 
\multicolumn{5}{c}
{\tablename\ \thetable\ -- \textit{Continued}} \\
\hline\hline
\noalign{\vspace{1mm}}
No & $v_{\text{rel}}$ [km s$^{-1}$] & Ion A & $\log N$(A) & $b$ [km s$^{-1}$]\\
\noalign{\vspace{1mm}}
\hline
\noalign{\vspace{1mm}}

8&$-10.2$	&	C\,{\sc ii}&	 $14.16\pm0.01$	&	$9.1\pm0.2$	\\ 
&	&	N\,{\sc ii}&	 $13.44\pm0.04(0.04)$	&	$6.4*$	\\ 
&	&	O\,{\sc i}&	 $14.15\pm0.00(0.03)$	&	$7.4\pm0.1$	\\ 
&	&	Mg\,{\sc i}&	 $11.47\pm0.02(0.12)$	&	$6.0\pm0.3$	\\ 
&	&	Mg\,{\sc ii}&	 $13.35\pm0.02$	&	$6.7\pm0.2$	\\ 
&	&	Al\,{\sc ii}&	 $12.23\pm0.01$	&	$6.2\pm0.1$	\\ 
&	&	Al\,{\sc iii}&	 $11.66\pm0.11(0.11)$	&	$10.7\pm3.5$	\\
&	&	Si\,{\sc ii}&	 $13.55\pm0.00$	&	$6.0\pm0.1$	\\ 
&	&	S\,{\sc ii}&	 $13.56\pm0.06$	&	$13.2\pm0.9$	\\ 
&	&	Fe\,{\sc ii}&	 $13.10\pm0.0(0.03)$	&	$5.3\pm0.1$	\\ \hline
\noalign{\vspace{1mm}}

9&15.4	&	C\,{\sc ii}&	 $13.58\pm0.01$	&	$10.2\pm0.3$	\\
&	&	N\,{\sc ii}&	 $-$	&	$-$	\\
&	&	O\,{\sc i}&	 $14.00\pm0.01(0.03)$	&	$14.8\pm0.3$	\\ 
&	&	Mg\,{\sc i}&	 $10.94\pm0.04(0.13)$	&	$6.0\pm0.7$	\\
&	&	Mg\,{\sc ii}&	 $12.66\pm0.02$	&	$9.5\pm0.5$	\\ 
&	&	Al\,{\sc ii}&	 $11.64\pm0.02$	&	$9.6\pm0.6$	\\ 
&	&	Al\,{\sc iii}&	 $-$	&	$-$	\\ 
&	&	Si\,{\sc ii}&	 $12.90\pm0.01$	&	$11.1\pm0.4$	\\ 
&	&	S\,{\sc ii}&	 $-$	&	$-$	\\ 
&	&	Fe\,{\sc ii}&	 $12.44\pm0.01(0.03)$	&	$5.0\pm0.2$	\\ \hline
\noalign{\vspace{1mm}}

10&26.4	&	C\,{\sc ii}&	 $12.75\pm0.05$	&	$3.1\pm0.2$	\\
&	&	N\,{\sc ii}&	 $-$	&	$-$	\\ 
&	&	O\,{\sc i}&	 $-$	&	$-$	\\ 
&	&	Mg\,{\sc i}&	 $10.87\pm0.08(0.14)$	&	$5.9\pm1.6$	\\
&	&	Mg\,{\sc ii}&	 $12.22\pm0.03$	&	$2.5\pm0.3$	\\ 
&	&	Al\,{\sc ii}&	 $10.90\pm0.12$	&	$2.2\pm0.8$	\\ 
&	&	Al\,{\sc iii}&	 $-$	&	$-$	\\ 
&	&	Si\,{\sc ii}&	 $12.21\pm0.04$	&	$2.3\pm0.3$	\\ 
&	&	S\,{\sc ii}&	 $-$	&	$-$	\\ 
&	&	Fe\,{\sc ii}&	 $12.30\pm0.01(0.03)$	&	$2.8\pm0.2$	\\ \hline
\noalign{\vspace{1mm}}

11&35.8	&	C\,{\sc ii}&	 $13.38\pm0.01$	&	$8.2\pm0.2$	\\ 
&	&	N\,{\sc ii}&	 $-$	&	$-$	\\ 
&	&	O\,{\sc i}&	 $13.43\pm0.02(0.04)$	&	$17.6\pm0.7$	\\
&	&	Mg\,{\sc i}&	 $10.40\pm0.18(0.22)$	&	$4.4\pm1.6$	\\
&	&	Mg\,{\sc ii}&	 $12.42\pm0.02$	&	$6.1\pm0.3$	\\ 
&	&	Al\,{\sc ii}&	 $11.51\pm0.03$	&	$7.0\pm0.5$	\\ 
&	&	Al\,{\sc iii}&	 $11.13\pm0.12(0.12)$	&	$7.3*$	\\
&	&	Si\,{\sc ii}&	 $12.44\pm0.02$	&	$6.6\pm0.3$	\\ 
&	&	S\,{\sc ii}&	 $-$	&	$-$	\\ 
&	&	Fe\,{\sc ii}&	 $11.90\pm0.03(0.04)$	&	$4.3\pm0.4$	\\ \hline
\end{tabular}
\tablefoot{
The fields marked with ``--'' indicate non-visible components which were excluded from the fit. The $b$-values marked by an asterisk have been fixed during the fitting procedure to achieve a successful solution. The additional error in the column densities caused by the fixing has been added quadratically to the column density error in each component of the concerning absorption line and the total column density error is given in brackets. All other errors are provided by \texttt{FITLYMAN}.
}
\end{table}

We find that a minimum of 11 individual components are required to optimally fit the absorption features evident in the weakly and 
highly ionised species with satisfying $\chi^2$-values. In the following, we number the individual velocity components consecutively, starting with the bluest absorption component. The absorber redshift is set at $z=1.839$. For the weakly ionised species, we identify three dominant velocity components at $-46.3$\,km\,s$^{-1}$, $-29.0$\,km\,s$^{-1}$, and $-10.2$\,km\,s$^{-1}$ (components 6,7, and 8). All subcomponents span a total velocity range of $\Delta v \approx200$\,km\,s$^{-1}$. The relative velocities of all subcomponents are listed in Table\ \ref{Fit results}.

The velocity components of the highly ionised species are less well separated. We identify four dominant velocity components 
at $-135.6$\,km\,s$^{-1}$, $-90.1$\,km\,s$^{-1}$, $-43.9$\,km\,s$^{-1}$, and $-6.1$\,km\,s$^{-1}$ (components 2,3,7, and 9). 
The relative velocities of all subcomponents are listed in Table\ \ref{Fit results C and Si 2}.

\subsection{Neutral and weakly ionised species}

In the H\,{\sc i} Lyman $\alpha$ absorption, the various velocity subcomponents are superposed in one extended Lyman trough. 
From a single-component fit of this trough, we obtain a total neutral hydrogen column density of $\log N($H\,{\sc i}$) = 19.48 \pm 0.01$, 
confirming the earlier results of \citet{Dessauges2003}. The centroid of the Lyman $\alpha$ line lies at $-36.2$\,km s$^{-1}$, thus between components 6 and 7 of the neutral gas components as traced by O\,{\sc i} (see Fig.\ \ref{fitovervelocityscale}). A component model of the H\,{\sc i} absorption, which is in line with the above given value for the total H\,{\sc i} column density, is presented in the Appendix. The fit results for the metal species are summarised in Table\ \ref{Fit results}. Unless stated otherwise, the column-density errors given in this paper are the $1\sigma$ fitting errors 
derived by \texttt{FITLYMAN}. They account for the statistical noise, but do not include uncertainties 
associated with the tracing of the continuum or systematic uncertainties arising from fixing of the line centres and $b$-values. 
We decide not to fit the Si\,{\sc iii} profiles because of severe saturation. Upper limits for weak or undetected absorption features are 
listed in Table\ \ref{Upper limit}. The comparison to Table\ \ref{Fit results} shows that 
for most of the weaker components the fitting procedure did not give reliable results. 
Some of the fitted components clearly are dominated by noise and the results should be interpreted with caution.

\begin{table}[h]
\caption{Detection limits \label{Upper limit}}
\centering
\scriptsize
\begin{tabular}{lc}\hline\hline
\noalign{\vspace{1mm}}
Ion A & Detection limit in $\log N$(A)  \\ 
\noalign{\vspace{1mm}}
\hline
\noalign{\vspace{1mm}}
O\,{\sc i} & 12.6 	\\
N\,{\sc i} & 12.6 	\\
N\,{\sc ii} & 13.3 	\\ 
Mg\,{\sc i} & 10.8 	\\
Al\,{\sc iii} & 11.7	\\
S\,{\sc ii} & 13.2 	\\
Fe\,{\sc ii} & 11.6	 \\  \hline
\end{tabular}
\end{table}

\subsection{Highly ionised species}

\begin{table}[htp]
\caption{Results of the fitting procedure for C\,{\sc iv} and Si\,{\sc iv}}
\centering
\scriptsize
\begin{tabular}{cccll}\hline\hline
\noalign{\vspace{1mm}}
No & $v_{\text{rel}}$ [km s$^{-1}$] & Ion A & $\log N$(A) & $b$ [km s$^{-1}$]  \\ 
\noalign{\vspace{1mm}}
\hline
\noalign{\vspace{1mm}}
1& $-$156.9	&    C\,{\sc iv}	& $13.03 \pm  0.01$ & $13.0 \pm  0.3$    \\
 &		&    Si\,{\sc iv}	&$12.66 \pm  0.00$ & $12.6 \pm  0.2$   \\
2& $-$135.7	&    C\,{\sc iv} 	& $13.31 \pm  0.01$ & $ 9.6 \pm  0.1$  \\
 &		&    Si\,{\sc iv}	& $12.98 \pm  0.00$ & $ 6.9 \pm  0.0$   \\
3& $-$92.7	&    C\,{\sc iv}	& $13.50 \pm  0.02$ & $20.6 \pm  0.5$   \\
 &		&   Si\,{\sc iv}	& $12.98 \pm  0.00$ & $20.9 \pm  0.3$   \\
4& $-$83.9	&    C\,{\sc iv}	& $12.72 \pm  0.06$ & $ 6.3 \pm  0.4$   \\
 &		&    Si\,{\sc iv}	& $12.77 \pm  0.01$ & $ 5.5 \pm  0.1$   \\
5& $-$70.9	&    C\,{\sc iv} 	& $12.71 \pm  0.09$ & $10.1 \pm  1.5$  \\
 &		&    Si\,{\sc iv}	& $12.29 \pm  0.01$ & $ 4.7 \pm  0.2$   \\
6& $-$53.2	&    C\,{\sc iv}	& $12.90 \pm  0.07$ & $10.0 \pm  0.8$   \\
 &		&    Si\,{\sc iv}	& $11.89 \pm  0.06$ & $ 6.1 \pm  0.7$   \\
7& $-$43.0	&    C\,{\sc iv}	&$13.28 \pm  0.03$ & $14.8 \pm  1.0$   \\
 &		&    Si\,{\sc iv} 	& $13.30 \pm  0.00$ & $13.3 \pm  0.1$  \\
8& $-$23.2	&    C\,{\sc iv}	& $12.49 \pm  0.07$ & $ 6.0 \pm  0.5$   \\
 &		&    Si\,{\sc iv}	& $12.34 \pm  0.02$ & $ 6.0 \pm  0.2$   \\
9& $-$6.1	&    C\,{\sc iv}	& $13.64 \pm  0.01$ & $10.1 \pm  0.2$   \\
 &		&    Si\,{\sc iv}	&$13.02 \pm  0.00$ & $11.5 \pm  0.1$   \\
10& 8.9		&    C\,{\sc iv}	& $13.03 \pm  0.04$ & $ 9.9 \pm  0.4$   \\
 &		&    Si\,{\sc iv}	& $12.12 \pm  0.02$ & $13.5 \pm  0.7$   \\
11& 36.2	&    C\,{\sc iv}	& $12.57 \pm  0.02$ & $15.5 \pm  0.7$   \\
 &		&    Si\,{\sc iv} 	& $12.35 \pm  0.01$ & $ 8.0 \pm  0.2$  \\ \hline
\noalign{\vspace{1mm}}
\end{tabular}
\tablefoot{For the multi-component Voigt-profile fitting procedure, we apply a velocity structure that resulted from a 
simultaneous fit of all absorption lines, but with individual Doppler parameters for each component 
of each ion. All errors are the errors provided by \texttt{FITLYMAN} and do not take into account systematic uncertainties. The velocity is given in exact values for C\,{\sc iv}. The velocities for the Si\,{\sc iv} components deviate by a maximum of $\pm 2.4$ km s$^{-1}$ 
because the line centres were allowed to be shifted slightly by \texttt{FITLYMAN} during the fitting procedure.
} \label{Fit results C and Si 2}
\end{table} 

Absorption by highly ionised gas is seen in C\,{\sc iv} $\lambda\lambda 1548, 1550$, and Si\,{\sc iv} $\lambda\lambda 1393,1402$. 
N\,{\sc v} and O\,{\sc vi} are located outside the observed wavelength range. We identify 11 absorption components in the high-ion absorption pattern. 
The results of the fitting procedure are summarised in Table\ \ref{Fit results C and Si 2}. We note that for the high ions the individual velocity components are less well separated than those of the weakly ionised species. This is evident for example in component 8, which seems to be mixed into component 9 in carbon, but in silicon it appears to be mixed into component 7 (see Fig.\ \ref{Vergleich}). Therefore, it is likely that
the true sub-component structure is not fully resolved.

\subsection{Comparison} \label{Comparison}

\begin{figure}
 \resizebox{\hsize}{!}{\includegraphics{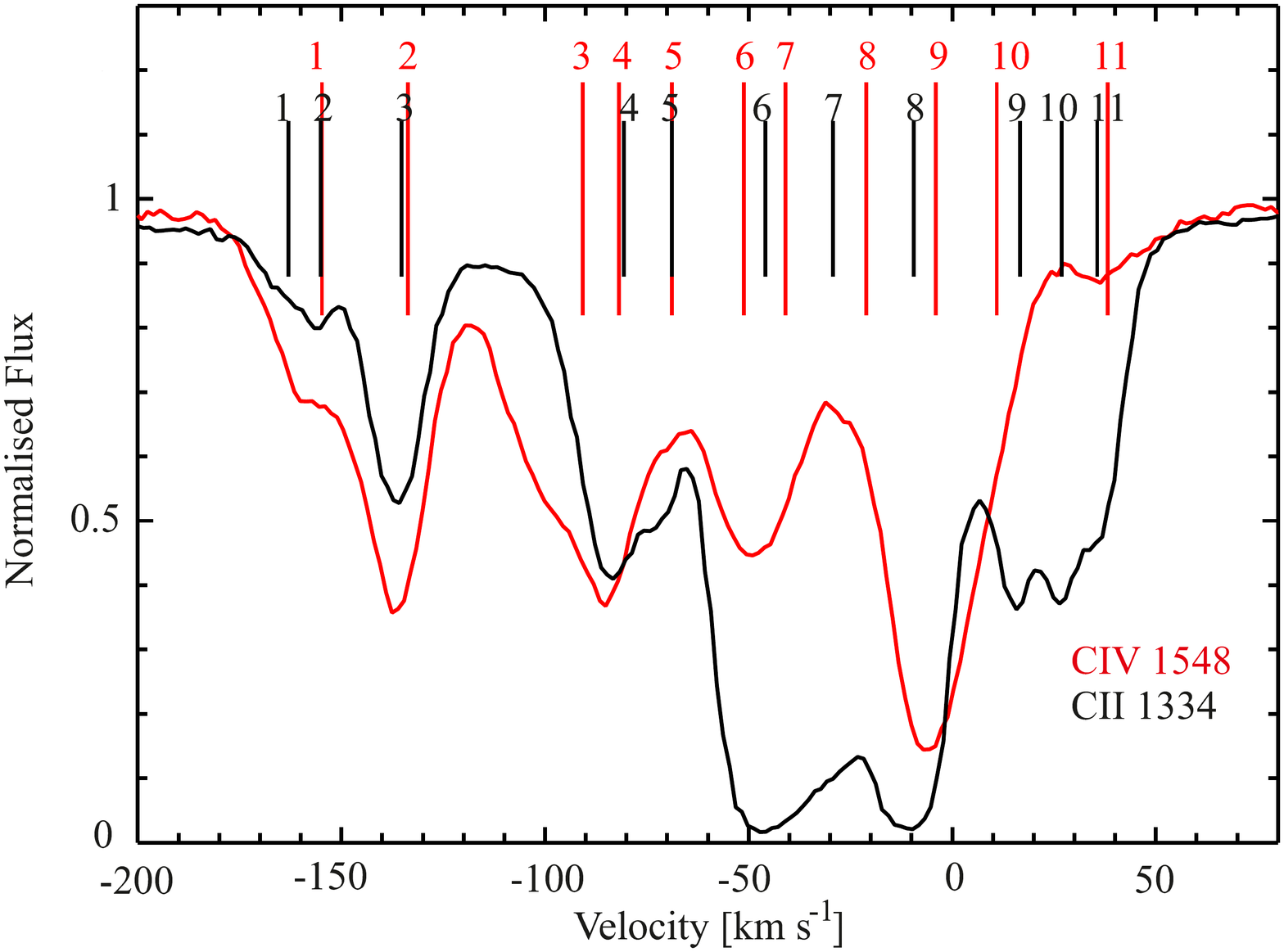}}
 \resizebox{\hsize}{!}{\includegraphics{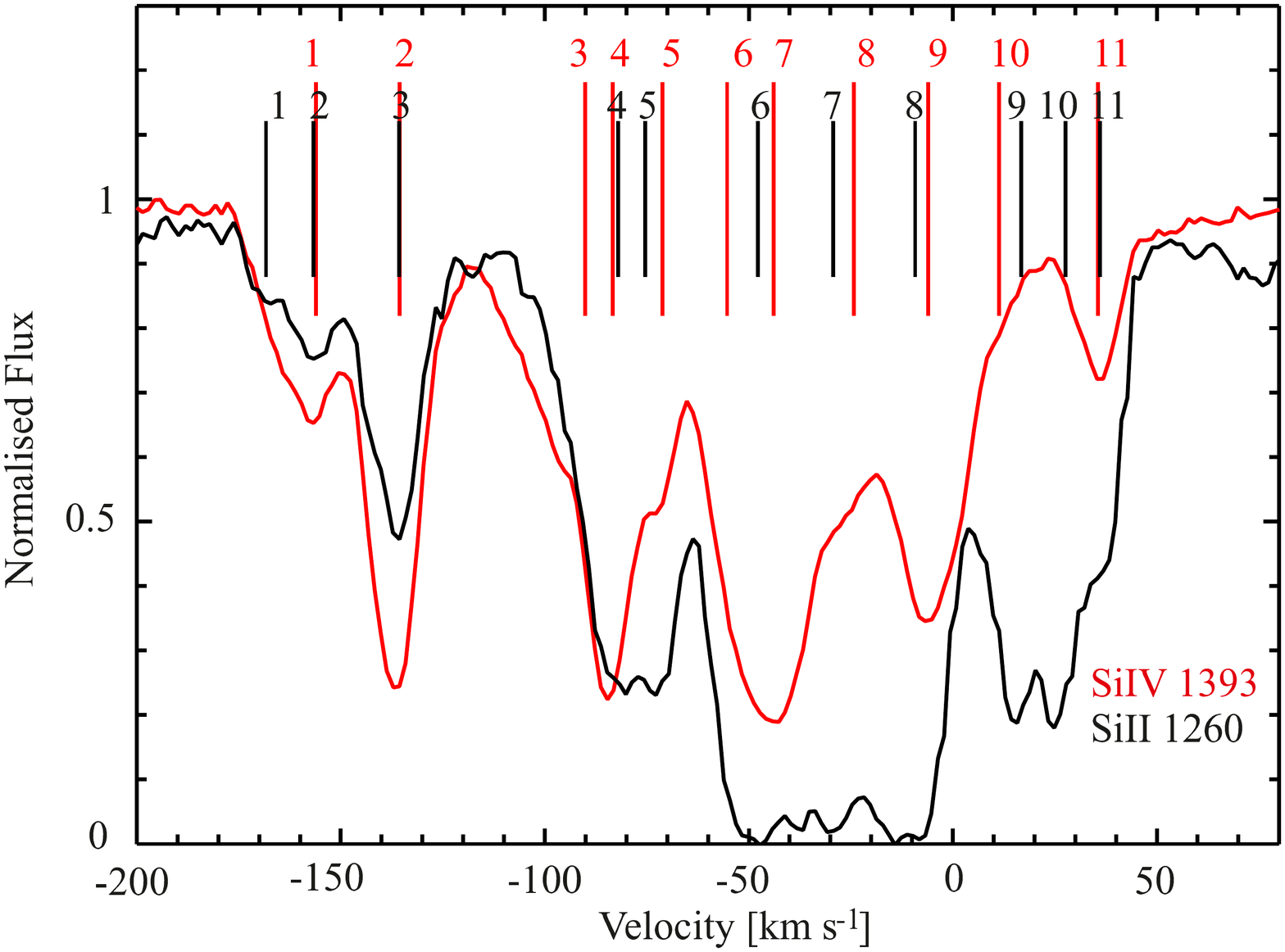}}
\caption{Comparison between the velocity structure of highly and weakly ionised species. Presented are C\,{\sc ii} (black) and C\,{\sc iv} (red) in the upper panel, and Si\,{\sc ii} (black) and Si\,{\sc iv} (red) in the lower panel. The vertical tick marks indicate the locations of the line centres of the absorption components.\label{Vergleich}}
\end{figure}

In Fig.\ \ref{Vergleich}, the structure of the absorption lines of the weakly and highly ionised species is compared. For this purpose, 
the velocity structure of the Mg\,{\sc ii} absorption line is used as reference for the weakly ionised species and applied to the 
C\,{\sc ii} and Si\,{\sc ii} absorption lines. In a similar way, the velocity structure that results from the simultaneous fit of 
C\,{\sc iv} and Si\,{\sc iv} is used as reference for the highly ionised species. 

The complex velocity structure in the different ions implies a multi-phase nature of the gas where the ionisation ratios vary among 
the different components. The gas seems to be more highly ionised in the four bluest subcomponents, whereas in the other subcomponents the weakly ionised species are more dominant. We also note that the line ratios between individual low-ion transitions vary among the different components. Component 5, for example, is apparently stronger in Si\,{\sc ii} while it is barely visible in C\,{\sc ii}.

\citet[][]{Dessauges2003} point out that in many sub-DLAs the high-ion transitions commonly show very different absorption profiles 
compared to low-ion transitions. This trend is also observed in DLAs \citep{Fox2011,Fox2007,Wolfe2005,Dessauges2001,Wolfe2000,Lu1996}.
In contrast to these previous findings, the velocity structure of the highly and weakly ionised species in the system 
towards QSO B1101$-$26 is rather similar. Most of the velocity components coincide within the uncertainty range for determining
the component's central velocity, which is of the order of 10 km\,s$^{-1}$. This value is calculated from the line centre 
errors given by \texttt{FITLYMAN}. Kinematic alignment between weakly and highly ionised species has also been found by some other studies \citep[e.g.][]{Lehner2008}. Especially the intermediate ion Al\,{\sc iii} has in general the same velocity distribution as the weakly ionised species \citep{Vladilo2001}.

\subsection{Abundances from ion ratios} \label{Abundances}

Throughout this work, abundances are given in the notation [X/Y]$=\log$($N$(X)/$N$(Y))$-\log($X/Y$)_{\odot}$ with solar abundances from \citet{Asplund2009}. In this notation, $N$(X) is defined as the sum of the column densities of all ionisation stages of element X. In practice, abundances are usually calculated on the basis of the column density of the dominant ionisation state of element X, because not all ionisation states are available and because it is assumed that the non-dominant ionisation states barely add to the total abundance of X. However, this might not always be true and therefore ionisation corrections will be performed in Sec.\ \ref{Cloudy} in order to compensate for significant amounts of X in different ionisation states. In order to distinguish between the real physical abundance ratios of the elements X and Y (or the ionisation corrected abundance) and the abundance ratio simply calculated by the dominant ionisation state, we use the terms [X/Y] and [A/B], respectively, where A is an ionisation state of element X and B is an ionisation state of element Y.

\begin{figure}
\resizebox{\hsize}{!}{\includegraphics{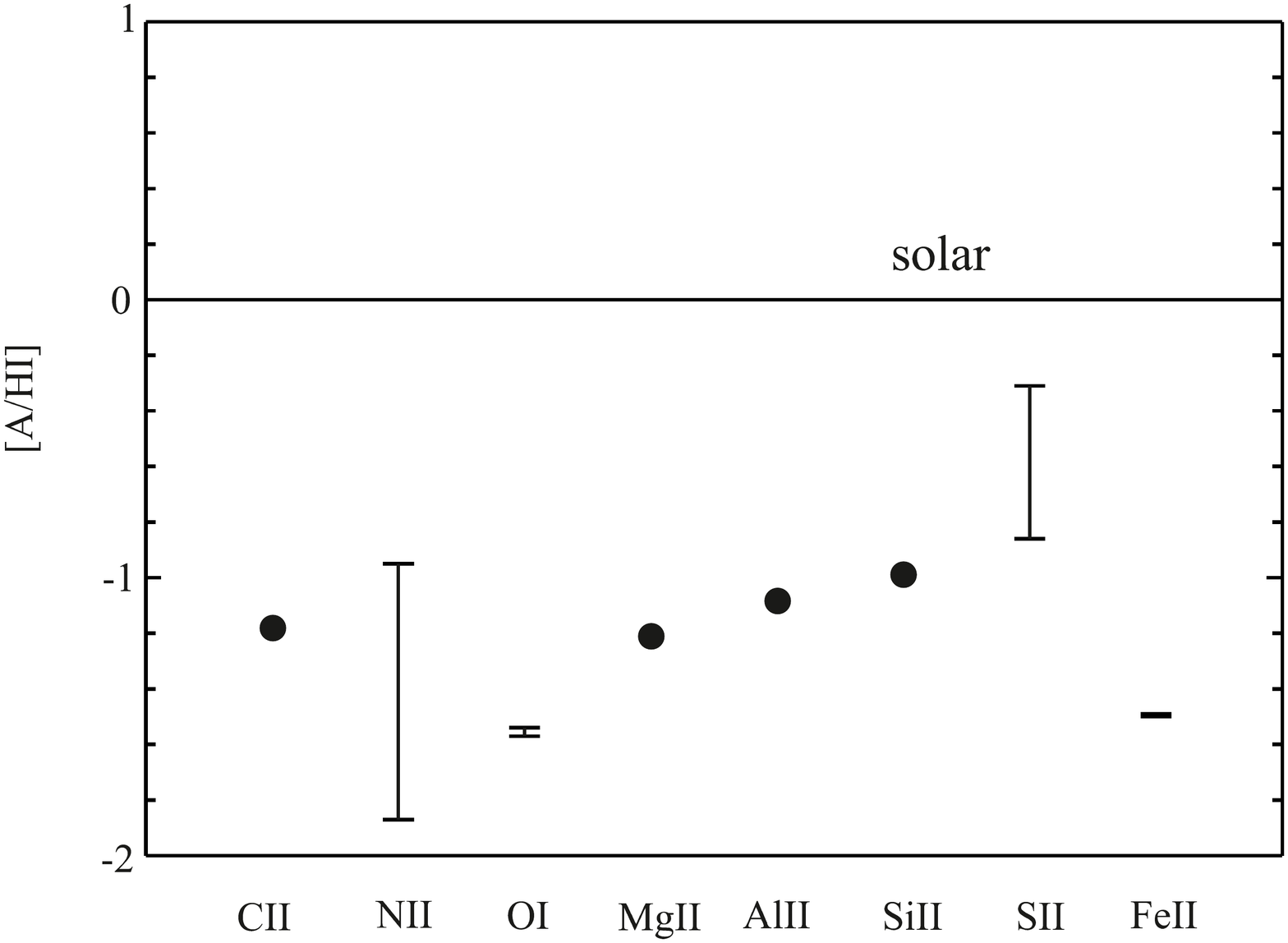}}
\caption{Comparison of element abundances in the absorption line system at $z=1.839$ and solar abundances. 
Exact values are marked by a filled circle with the error bars being smaller than the symbols itself.
 Upper and lower limits are marked by horizontal bars. \label{Element abundances figure}}
\end{figure}

First, we discuss the results for the conventional approach without ionisation corrections. In Table\ \ref{Element abundances table} and Fig.\ \ref{Element abundances figure}, the total column densities for the detected species (derived from summing over all absorption components) and the integrated metal abundances are shown. These values have been derived by taking into account the detection limits presented in Table\ \ref{Upper limit}, i.e. measured column densities that lie below the formal detection limit have not been considered. The overall metallicity of the absorber, as traced by the $\alpha$-element oxygen, 
is [O\,{\sc i}/H\,{\sc i}] $= -1.56 \pm 0.01$, and is therefore higher than what has been found by \citet{Dessauges2003} who estimate [O/H] $= -1.78 \pm 0.12$.  However, from Fig.\ \ref{Element abundances figure} and Table\ \ref{Element abundances table} O\,{\sc i} appears to be underabundant compared to all other available ions. A plausible reason for this apparent underabundance could be an ionisation effect, which will be  examined further as part of 
the Cloudy modelling presented in Sec.\ \ref{Cloudy}. \citet{Dessauges2003} also show that S\,{\sc ii} appears to be strongly overabundant compared to the other $\alpha$-elements, especially compared to oxygen. We note that the sulphur abundance obtained here deviates substantially (by 0.73\,dex) from the sulphur abundance obtained by \citet{Dessauges2003}. However, the dominant contribution to this discrepancy comes from different solar reference abundances 
used in these two studies (\citet{Grevesse1998} vs. \citet{Asplund2009}).
There is no obvious origin for such an overabundance, as sulphur and oxygen are jointly produced in the $\alpha$-process and are 
both non-refractory elements, i.e. they are not influenced by dust-depletion \citep[e.g.][]{Jenkins2009}.

Nevertheless, the ionisation potential of S\,{\sc ii} is 23.23\,eV, i.e. it is significantly higher than those of O\,{\sc i} 
and H\,{\sc i} ($\sim 13.6$\,eV for both). This implies that S\,{\sc ii} still exists in gas where part of the hydrogen is ionised. By contrast, neutral oxygen is coupled to hydrogen by strong charge-exchange reactions \citep{Osterbrock2006,Field1971}. As a consequence, the fraction of oxygen and hydrogen in ionised form should be identical, and if the larger part of the hydrogen is ionised, so will be the oxygen. This aspect will be  discussed further in Sec.\ \ref{Cloudy}.

We note that in our discussion we consider the {\it mean} abundances in this system, averaged over several components unless stated otherwise. 
However, abundance ratios in individual components may significantly deviate from the mean value. This is apparent from 
Fig.\ \ref{Vergleich}. Such behaviour has also been found in other systems, see for example \citet{Ledoux2003} and 
\citet{Richter2005}. In the following, particularly interesting abundance ratios are discussed.

\begin{table*}
\caption{ Element abundances in the absorption line system at $z=1.839$.}
\centering
 \label{Element abundances table}
\begin{tabular}{lllllll}\hline\hline
\noalign{\vspace{1mm}}
Ion A & $\log N$(A)$_{\text{tot}}$ & Element X& $\log N$(X)$_{\odot}$&$\log$(A/H\,{\sc i}) & [A/H\,{\sc i}] & [X/H]$^{2}$\\ 
\noalign{\vspace{1mm}}
\hline
\noalign{\vspace{1mm}}
C\,{\sc ii} & $14.73 \pm 0.01$ 	& C &$8.43 \pm 0.05$ &$-4.75 \pm 0.01$ & $-1.18 \pm 0.05$	&$-0.96\pm0.10$\\
N\,{\sc i} & $\leq 12.6 $ & N & $7.83 \pm 0.05$ & $\leq -6.88 $ &$\leq -2.71 $& $\ldots$ \\
N\,{\sc ii} & $13.44 \ldots 14.36^{1}$	& N & $7.83 \pm 0.05$ &$-5.12 \ldots -6.04^{1}$& $-0.95 \ldots -1.87^{1}$	&$\ldots$\\
O\,{\sc i} & $14.60 \ldots 14.62^{1}$ 	& O & $8.69 \pm 0.05$ &$-4.86 \ldots -4.88^{1}$&$ -1.55 \ldots -1.57^{1}$	&$-1.78\pm0.12$\\
Mg\,{\sc ii} & $13.87 \pm 0.01$	& Mg &$7.60 \pm 0.04$ & $-5.61 \pm 0.01$ & $-1.21 \pm 0.04$ 	&$-1.01\pm0.06$ \\ 
Al\,{\sc ii} & $12.85 \pm 0.01$	& Al &$6.45 \pm 0.03$ &$-6.63 \pm 0.01$ & $-1.08 \pm 0.03$	&$-1.14\pm0.08$\\
Al\,{\sc iii} & $11.96 \ldots 12.77^{1}$ & Al &$6.45 \pm 0.03$& $-7.52 \ldots -6.71^{1}$& $-1.97 \ldots -1.16^{1}$& $\ldots$\\
Si\,{\sc ii} & $14.00 \pm 0.01$	& Si &$7.51 \pm 0.03$ &$-5.48 \pm 0.01$ & $-0.99 \pm 0.03$	&$-1.06\pm0.05$\\
S\,{\sc ii} & $13.74 \ldots 14.29^{1}$	& S & $7.12 \pm 0.03$ &$-5.19 \ldots -5.74^{1}$& $-0.31 \ldots -0.86^{1}$	&$-1.04\pm0.13$\\
Fe\,{\sc ii} & $13.48 \ldots 13.49^{1}$ & Fe & $7.50 \pm 0.04$ &$-5.99 \ldots -6.00^{1}$& $-1.49 \ldots -1.5^{1}$	&$-1.49\pm0.05$\\ \hline
\end{tabular}
\tablefoot{$^{1}$column density range takes into account the derived upper limits for individual subcomponents. 
$^{2}$Reference: \cite{Dessauges2003}}
\end{table*}

\textit{Nitrogen} -- The [N/O] ratio provides important information on the chemical evolution and the enrichment
history of an absorption system. In the absorber considered here, N\,{\sc i} is not detected, but we derive a useful upper 
limit for N\,{\sc i} in the main component (8) (see Table\ \ref{Upper limit}).
From this, we also obtain an upper limit for  the nitrogen-to-oxygen-ratio of [N\,{\sc i}/O\,{\sc i}]$_{\text{8}} \le -0.65$ or $\log$(N\,{\sc i}/O\,{\sc i})$_{\text{8}} \le -1.55$. This low N\,{\sc i}/O\,{\sc i}  ratio indicates a chemically young system with a delayed production of nitrogen in intermediate-mass stars compared to the quick production of $\alpha$-elements in massive stars. This delay appears to be even more pronounced if 
one considers the apparent sulphur abundance, yielding [N\,{\sc i}/S\,{\sc ii}]$_{\text{8}} \le -2.26$. 
Similarly low N/$\alpha$ ratios have also been found in other studies \citep[e.g.][]{Prochaska2002b}. For example, \cite{Pettini2008} have found 
$\log$(N/O$)=-2.53$ to $-1.14$ in their study on metal-poor DLAs. A comparison to their findings about the primary or 
secondary origin of nitrogen places the absorption system considered here close to the primary plateau. However, ionisation effects 
can considerably affect the calculation of the nitrogen abundance in sub-DLAs because of the large photoionisation-cross 
section of neutral nitrogen \citep{Sofia1998}. This aspect will be examined further in Sec.\ \ref{Cloudy}.

\textit{$\alpha$-elements and iron} -- An overabundance of $\alpha$-elements compared to iron is usually interpreted as 
the result of the chemical enrichment of protogalactic structures by SN Type II explosions. That is because iron is predominantly produced in SN Type Ia explosions \citep{Matteucci1986,Acharova2013,Yates2013} and the progenitors of the SN Type Ia have much longer lifetimes than the progenitors of SNe Type II. A high $\alpha$/Fe ratio is observed in nearly all DLAs \citep[e.g.][]{Prochaska2002} and also in metal-poor stars of our Galactic Halo, which must have preserved the abundance pattern of the young Milky Way \citep[e.g.][]{Nissen2004,Suda2011,Rafelski2012}. In the case of absorption line systems, the conclusion is less straight-forward, because the lack of iron in the gas can also be attributed to dust-depletion. Iron is strongly depleted onto dust grains in the local interstellar medium \citep[][]{SavageSembach1996} and therefore is expected to be underabundant in gas that contains interstellar dust.

In this study, we obtain

\begin{align*}
 1.11 \le \log \left( \frac{\text{O\,{\sc i}}}{\text{Fe\,{\sc ii}}} \right) & \le 1.14 \text{, or } -0.08 \le \text{[O\,{\sc i}/Fe\,{\sc ii}] } \le -0.05\\
 0.51 \leq \log \left( \frac{\text{Si\,{\sc ii}}}{\text{Fe\,{\sc ii}}} \right) & \le 0.52 \text{, or }0.50 \le \text{[Si\,{\sc ii}/Fe\,{\sc ii}] } \le 0.51 \text{, and}\\
 0.25 \le \log \left( \frac{\text{S\,{\sc ii}}}{\text{Fe\,{\sc ii}}} \right)  & \le 0.81 \text{, or } 0.63 \le [\text{S\,{\sc ii}/Fe\,{\sc ii}}] \le 1.19. 
\end{align*}

The values for silicon and sulphur suggest an overabundance of $\alpha$-elements compared to iron, even though silicon is expected 
to be affected by dust-depletion, too \citep{SavageSembach1996}. In contrast, the oxygen-to-iron ratio is 
significantly lower than other $\alpha$/Fe ratios in sub-DLAs and DLAs and would thus imply a chemically much more evolved 
system, which seems unlikely in view of the low N\,{\sc i}/$\alpha$ ratio and the considered redshift. 
This discrepancy has its origin in the largely inconsistent $\alpha$-element abundances in the absorber:

\begin{align*}
 0.69 \le \text{[S\,{\sc ii}/O\,{\sc i}]}  &\le 1.26\\
0.13 \le \text{[S\,{\sc ii}/Si\,{\sc ii}] } &\le 0.68\\
[\text{Si\,{\sc ii}/O\,{\sc i}}] &= 0.57 \pm 0.07.
\end{align*}

These ratios do not agree with current nucleosynthetic models in which all $\alpha$-elements are produced at the same time 
by massive stars, as supported by the observed relative solar abundances of $\alpha$-elements in Galactic halo stars \citep{Nissen2004}. 
However, similar inconsistencies in the relative abundance of $\alpha$-elements have been found in other
high-redshift absorbers, too \citep[e.g.][]{Bonifacio2001,Prochaska2002,Fathivavsari2013}.

We note that the large range in the above listed ratios originates in the very weak sulphur absorption, which is below the detection limit
in most of the absorption components. A closer look at component 8 reveals a trend possibly caused by ionisation effects that 
could be the reason for the inconsistent $\alpha$ abundances: [S\,{\sc ii}/O\,{\sc i}]$_{\text{8}}=0.98 \pm 0.09$, [Si\,{\sc ii}/O\,{\sc i}]$_{\text{8}}=0.58 \pm 0.07$, [S\,{\sc ii}/Si\,{\sc ii}]$_{\text{8}}=0.4 \pm 0.07$. In this component, sulphur seems to be the most abundant $\alpha$-element, followed by silicon, and oxygen. This sequence reflects the decreasing order of ionisation potentials of these ions, which are $23.23$\,eV, 16.34\,eV, and 13.61\,eV 
for S\,{\sc ii}, Si\,{\sc ii}, and O\,{\sc i}, respectively, suggesting that ionisation could be determining the relative strength
of the column densities of these ions (see Sec.\ \ref{Cloudy}).

\textit{Carbon} -- Previous studies of intervening absorbers with achievable carbon measurements have observed an underabundance of carbon 
compared to $\alpha$-elements \citep{Fechner2009,Pettini2008,Erni2006,Richter2005}. This is expected from modelling \citep[e.g.][]{Yates2013} and comparison to metal-poor halo stars \citep{Fabbian2009,Pettini2008}. Here, we obtain [C\,{\sc ii}/Fe\,{\sc ii}] $=0.32\pm0.06$, [C\,{\sc ii}/O\,{\sc i}] $=0.38\pm0.07$, $-0.87\pm 0.06 \le$[C\,{\sc ii}/S\,{\sc ii}] $\le 0.32\pm0.06$, [C\,{\sc ii}/Si\,{\sc ii}] $=-0.19\pm0.06$, and [C\,{\sc ii}/Al\,{\sc ii}$]=-0.10\pm0.06$. Hence, an underabundance of carbon can only be confirmed compared to silicon, sulphur, and aluminium. We note, however, that the carbon absorption line is blended with a weak absorption feature related to an absorption line system at $z=1.268$, for which a transition of Al\,{\sc ii} coincides with the carbon line at $z=1.839$. 

\textit{Odd-even effect} -- Elements with even atomic numbers are expected to have higher relative abundances because of the 
strong coupling of pairs of nucleons \citep{Arnett1971}. This effect is probed by the elements Si and Al, with Al being 
the odd element. \citet{Prochaska2002} have found enhanced Si/Al for the majority of the observed DLA systems, even 
though the ratios typically are lower than those of metal-poor stars at a similar metallicity \citep{McWilliam1995}. 
The measured values for our system ([Si\,{\sc ii}/Al\,{\sc ii}] $=0.09 \pm 0.06$) do not indicate a significant odd-even abundance effect.

\textit{Dust depletion} -- It is well known that in Galactic interstellar dust grains the elements iron, aluminium, and magnesium are 
refractory, silicon and carbon are mildly refractory, and sulphur is essentially non-refractory \citep{SavageSembach1996}.
Therefore, it is  interesting to examine the abundance ratios of refractory and non-refractory elements. We find

\begin{align*}
-0.87 \leq& \text{ [C\,{\sc ii}/S\,{\sc ii}]} \leq -0.32 \\
-0.90 \leq& \text{ [Mg\,{\sc ii}/S\,{\sc ii}]} \leq -0.35 \\
-0.77 \leq& \text{ [Al\,{\sc ii}/S\,{\sc ii}]} \leq -0.22 \\
-0.68 \leq& \text{ [Si\,{\sc ii}/S\,{\sc ii}]} \leq -0.13 \\
\text{and} -0.63 \leq& \text{ [Fe\,{\sc ii}/S\,{\sc ii}]} \leq -1.19.
\end{align*}

Albeit the large errors, these values are broadly consistent with depletion values for warm Galactic disk gas and halo gas 
\citep{SavageSembach1996,Weltly1999}. Nevertheless, the effects of dust depletion are likely entangled with nucleosynthetic 
and ionisation effects in this system, as will be discussed below.

\subsection{Photoionisation modelling} \label{Cloudy}

\begin{figure*}
 \noindent
\begin{minipage}[t]{.46\linewidth}
\includegraphics[width=\linewidth]{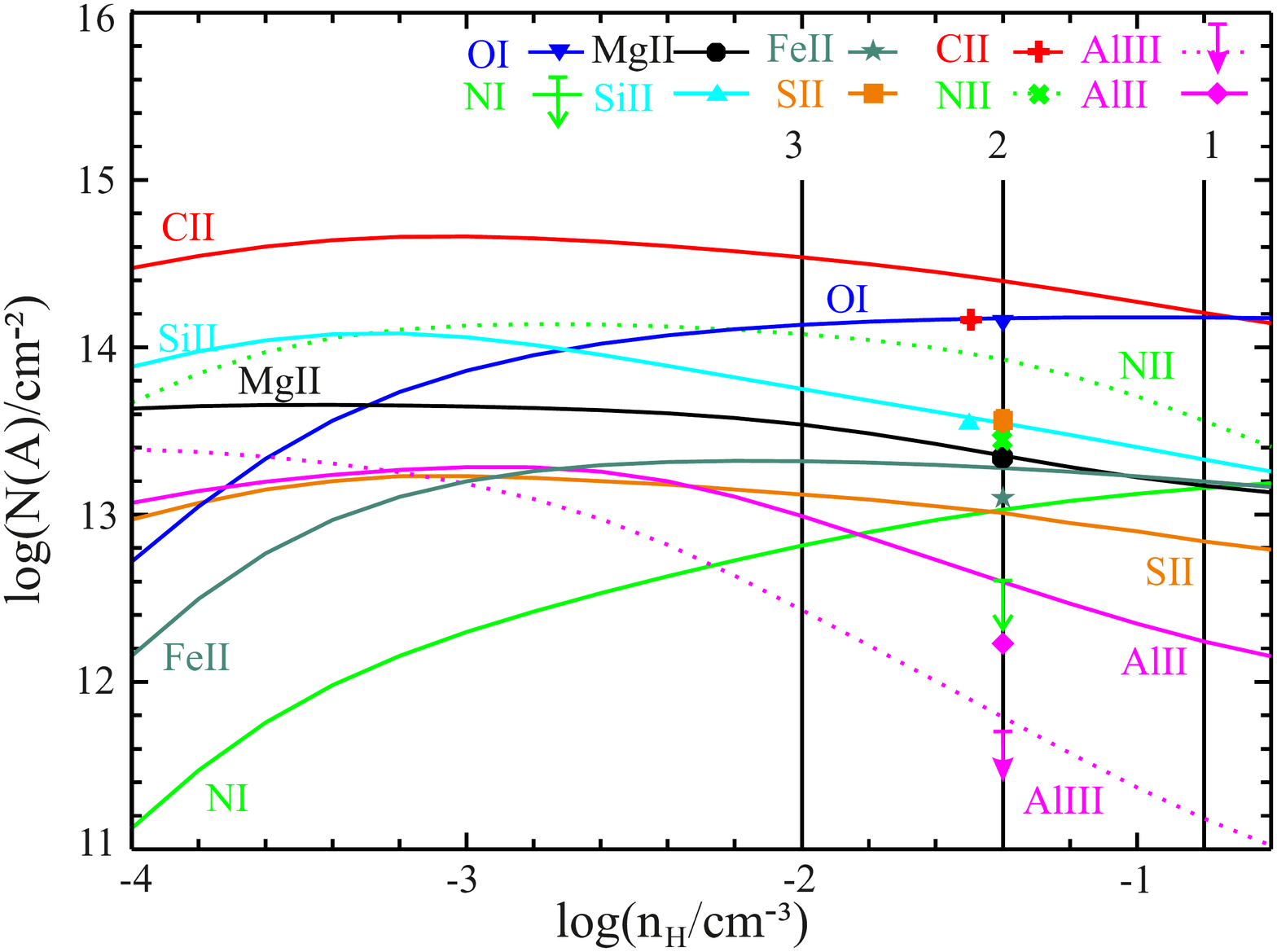}  
a) Component 8, metallicity of $2.75\%$ solar. The vertical black lines indicate the models 8.1, 8.2, and 8.3 with decreasing hydrogen density.
\end{minipage} \hfill
\begin{minipage}[t]{.46\linewidth}
 \includegraphics[width=\linewidth]{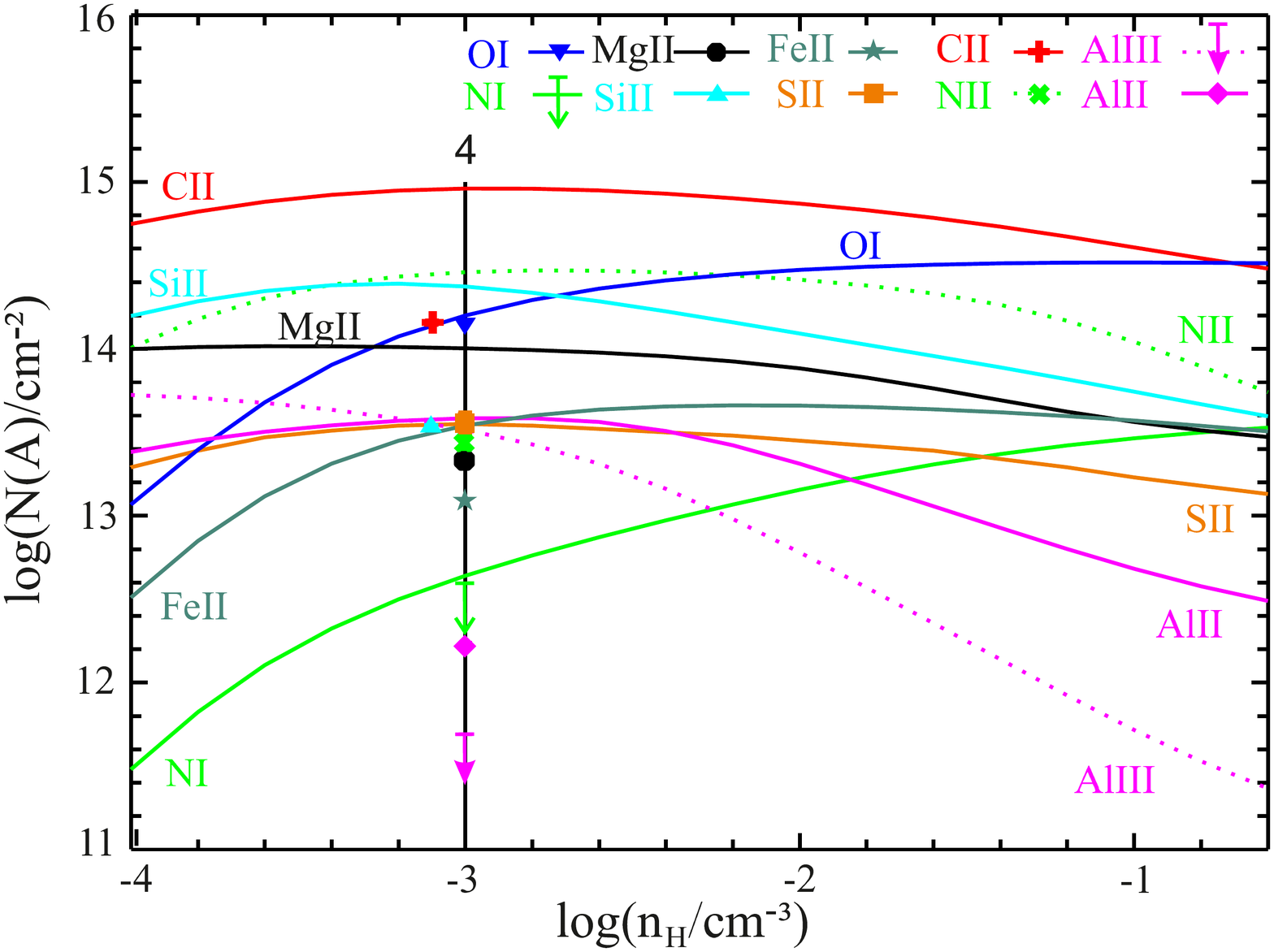}  
b) Component 8, metallicity of $6\%$ solar. The vertical black line indicates the model 8.4.\end{minipage} 

\begin{minipage}[t]{.46\linewidth}
\includegraphics[width=\linewidth]{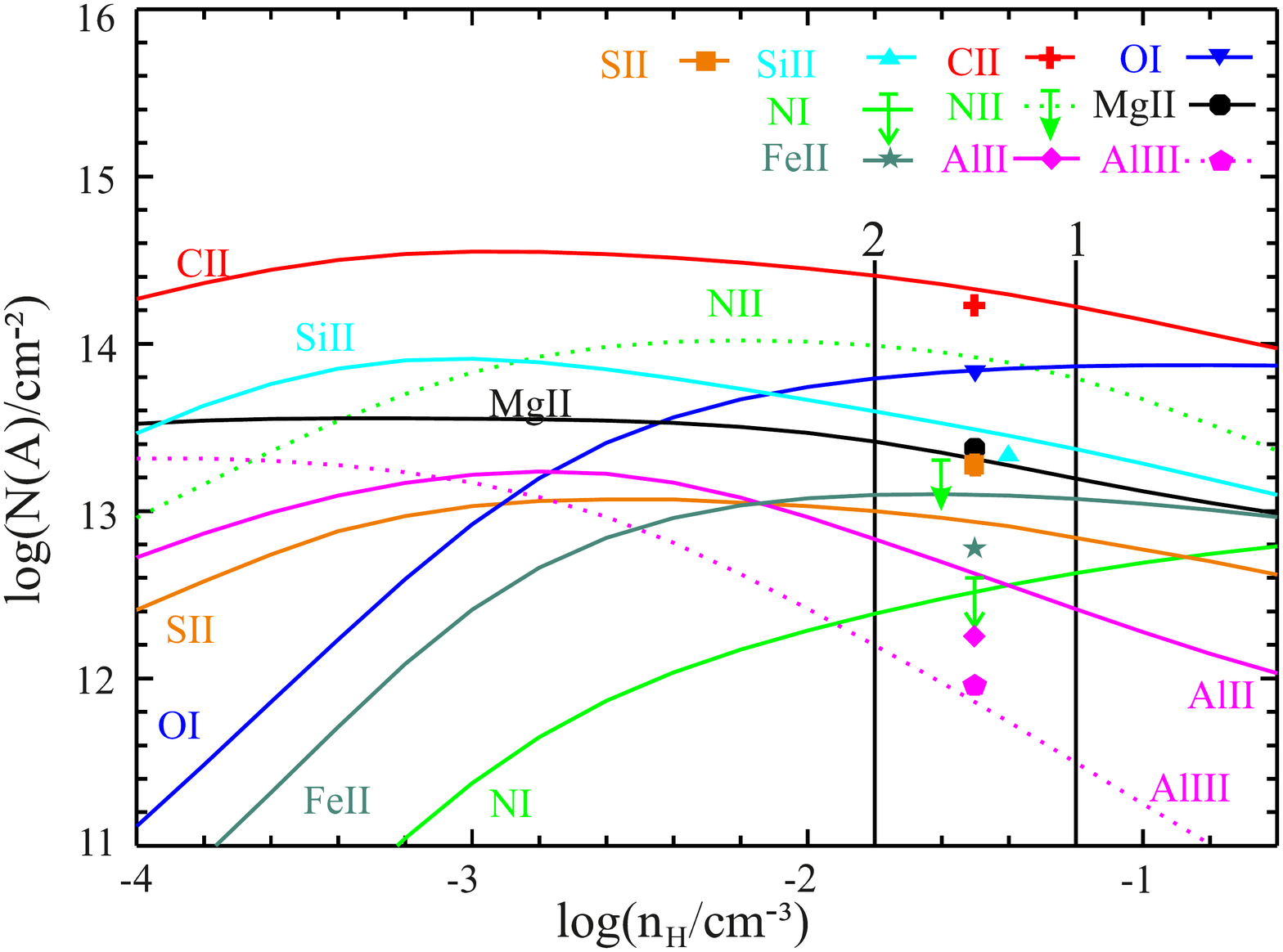} 
c) component 6, metallicity of 2.75\% solar. The vertical black lines indicate the models 6.1, and 6.2 with decreasing hydrogen density.
\end{minipage} \hfill
\begin{minipage}[t]{.46\linewidth}
\includegraphics[width=\linewidth]{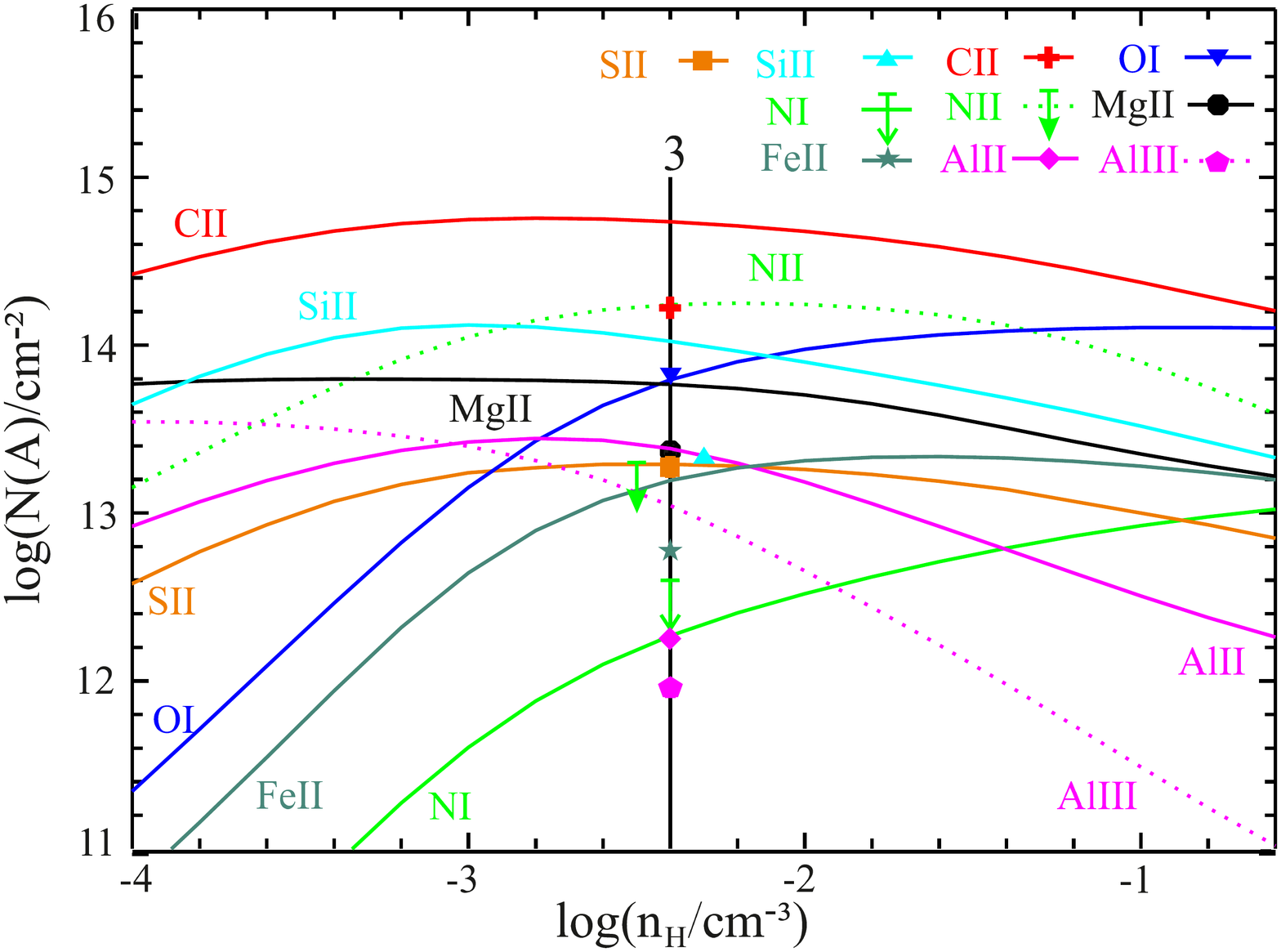}
d) component 6, metallicity  of 4.7\% solar. The vertical black line indicates the model 6.3.
\end{minipage} 
\caption{Cloudy models for component 8 (a,b) and 6 (c,d) in the absorption line system at $z=1.839$. 
The lines represent the Cloudy predictions and the symbols the measured values. The correlation between the symbols 
and the lines is shown in the legends in the upper right corners. Circles correspond to measured values 
with the error bars being smaller than the symbol itself except for sulphur. For Al\,{\sc iii} and N\,{\sc i} an upper limit is given in a) and b). The vertical black lines indicate the hydrogen density of the specific models (see text). Some symbols have a horizontal offset for clarity. \label{Cloudy plots} }
\end{figure*}

\begin{table*}
\begin{minipage}{.46\linewidth}
 
\caption{Deviations between measured column densities and Cloudy calculations in component 8}
\centering
\begin{tabular}{lcccc}\hline\hline
\noalign{\vspace{1mm}}
Ion A & \multicolumn{4}{c}{$\Delta \log N(A)$} \\
&  Model 8.1 &  Model 8.2 &  Model 8.3 &  Model 8.4 \\ 
\noalign{\vspace{1mm}}
\hline
\noalign{\vspace{1mm}}
C\,{\sc ii}	& 0.05 & 0.24 & 0.38 &  0.8	 \\ 
N\,{\sc ii}	& 0.12 & 0.49 & 0.64 & 1.02	\\ 
O\,{\sc i}	& 0.03 & 0.02 & $-0.02$ & 0.05	\\ 
Mg\,{\sc ii}	& $-0.18$ & 0 & 0.19 & 0.65	\\ 
Al\,{\sc ii}	& 0.01 & 0.37 & 0.76 & 1.35	\\
Al\,{\sc iii} 	& $\geq -0.52$ & $\geq$ 0.09 & $\geq$ 0.73 & $\geq$ 1.82	\\
Si\,{\sc ii} 	& $-0.22$ & 0 & 0.20 & 0.82 	\\ 
Fe\,{\sc ii} 	& 0.10 & 0.18 & 0.22 &  0.44	\\
S\,{\sc ii} 	& $-0.71$ & $-0.55$ & $-0.43$ & $-0.01$	\\ \hline
\noalign{\vspace{1mm}}
\multicolumn{5}{l}{Are the ion ratios reproduced correctly?} \\
(Al\,{\sc ii}/Al\,{\sc iii})& no & yes & yes & no	\\ 
(N\,{\sc i}/N\,{\sc ii})& no & yes & yes & yes	\\ \hline
\noalign{\vspace{1mm}}
\end{tabular}
 \label{Deviations component 8}
\end{minipage} \hfill
\begin{minipage}{.46\linewidth}
\caption{Deviations between measured column densities and Cloudy calculations in component 6}
\centering
\begin{tabular}{lccc}\hline\hline
\noalign{\vspace{1mm}}
Ion A & \multicolumn{3}{c}{$\Delta \log N(A)$} \\
&  Model 6.1 &  Model 6.2 &  Model 6.3  \\ 
\noalign{\vspace{1mm}}
\hline
\noalign{\vspace{1mm}}
C\,{\sc ii}	& 0 & 0.19 & 0.52 	 \\ 
N\,{\sc ii}	& 0.65 & 0.85 & 1.10 	\\ 
O\,{\sc i}	& 0.04 & $-0.04$ & $-0.04$ 	\\ 
Mg\,{\sc ii}	& $-0.18$ & 0.05 & 0.40 	\\ 
Al\,{\sc ii}	& 0.16 & 0.58 & 1.13 	\\
Al\,{\sc iii} 	& $ -0.46$ & 0.24 & 1.09 	\\
Si\,{\sc ii} 	& 0.04 & 0.27 & 0.69  	\\ 
Fe\,{\sc ii} 	& 0.29 & 0.31 & 0.41 	\\
S\,{\sc ii} 	& $-0.43$ & $-0.27$ & 0.01 	\\ \hline
\noalign{\vspace{1mm}}
\multicolumn{4}{l}{Are the ion ratios reproduced correctly?} \\
(Al\,{\sc ii}/Al\,{\sc iii})& no & no & yes	\\ 
(N\,{\sc i}/N\,{\sc ii})& yes & yes & yes 	\\ \hline
\noalign{\vspace{1mm}}
\end{tabular}
 \label{Deviations component 6}
\end{minipage}
\end{table*}

Intervening absorption line systems with large H\,{\sc i} column densities beyond a few times $10^{19}$ cm$^{-2}$ are usually assumed to 
be predominantly neutral because of self-shielding of the gas. For such systems, the metallicity in the neutral gas phase 
is usually determined by comparing the column densities of the dominant (low) ions of heavy elements with the column density of 
neutral hydrogen, as presented in Sec.\ \ref{Abundances}. While the approach is well justified for DLAs \citep[e.g.][]{Wolfe2005,Vladilo2001,Viegas1995} and many sub-DLAs \citep[e.g.][]{Prochaska2006,Peroux2006,Dessauges2003}, ionisation effects may become important for a specific class of sub-DLAs, namely those systems, in which the gas density, and thus the recombination rate, is relatively low, so that the neutral gas fraction falls 
substantially below unity. These ionisation effects could at least qualitatively explain the unexpected abundances in the previous section. Interestingly, \citet{Milutinovic2010} also report on super-solar metallicities ([S\,{\sc ii}/H\,{\sc i}]$=+0.36$) in a multi-phase sub-DLA \textit{before} applying ionisation corrections.

To estimate the influence of ionisation on the determination of metal abundances in our sub-DLA, we carried out photoionisation models using 
version 10.00 of the Cloudy photoionisation code \citep{Ferland1998}. For this purpose, we assume the incident ionising radiation field to be 
composed of the cosmic background radiation and a Haardt \& Madau (2001) UV background as implemented in Cloudy, both evaluated at the redshift of the absorber. 
We modelled a plane-parallel slab of gas and varied the hydrogen density $n_{\text{H}}$ between $10^{-5.0}$ and $10^{-0.6}$\,cm$^{-3}$ in steps of $10^{-0.2}$\,cm$^{-3}$. The temperature was left as a free parameter. Nevertheless, Cloudy adopted an electron temperature of about $T_{e} \approx 1-2 \cdot 10^{4}$\,K in every case. Because of the multi-phase nature of the sub-DLA with its different velocity components and a non-uniform gas density among the different components (resulting in a non-uniform ionisation parameter), we set up photoionisation models solely for selected velocity components, for which sufficient spectral line diagnostics are available. We have estimated the H\,{\sc i} column density in each component on the basis of the observed column density of O\,{\sc i} in the respective component and the assumption that the (O\,{\sc i}/H\,{\sc i}) ratio is constant throughout the absorber. This assumption is well justified, because of the strong connection of O\,{\sc i} and H\,{\sc i} by charge-exchange reactions (see also Fig.\ \ref{ModelHIauskonstOH}). The simulations were stopped once this H\,{\sc i} column density was reached.
In our initial modelling approach, we assume a solar chemical abundance pattern taken from \citet{Asplund2009} and scale down the 
overall metallicity in the gas to 2.75\% solar which corresponds to the measured [O\,{\sc i}/H\,{\sc i}]. Subsequently, we modify the overall metallicity and gas density in order to reproduce the observations, i.e. to best match the column density results obtained by \texttt{FITLYMAN}.

Fig.\ \ref{Cloudy plots}a) and b) show the Cloudy models for component 8. The H\,{\sc i} column density for this 
component is estimated to be $\log N(\text{H\,{\sc i}})= 19.02$. We find four models that partly match the observations. The deviations from the measured values are presented in Table\ \ref{Deviations component 8}. In the following, we will discuss their advantages and drawbacks.

\textit{Model 8.1} -- Model 1 is based on a metallicity of 2.75\% solar metallicity, which corresponds to the measured value of [O\,{\sc i}/H\,{\sc i}], and a hydrogen density of $\log n_{\text{H}}= -0.8$. This model shows the mathematically smallest deviations from the measured values.  However, it has some significant drawbacks. O\,{\sc i} is slightly overproduced at this density. By contrast, S\,{\sc ii} is overabundant compared to the Cloudy calculations. Mg\,{\sc ii} is slightly overabundant as well and the ratios of Al\,{\sc ii}/ Al\,{\sc iii} and N\,{\sc i}/ N\,{\sc ii} can not be reproduced.  Besides, the high hydrogen density seems unlikely for an intergalactic sub-DLA. However, the Al\,{\sc ii}/ Al\,{\sc iii} ratio might suffer from uncertainties in the recombination rates \citep{Vladilo2001}.

\textit{Model 8.2} -- This model assumes again a metallicity of 2.75\% solar but a hydrogen density of $\log n_{\text{H}}= -1.4$. It produces the measured column densities of O\,{\sc i}, Si\,{\sc ii}, and Mg\,{\sc ii} correctly, i.e. all available $\alpha$-elements except for sulphur which is again overabundant compared to the model. However, the Cloudy model predicts too much C\,{\sc ii}, Fe\,{\sc ii}, aluminium, and nitrogen. An underabundance of Fe\,{\sc ii} and nitrogen in this system could be due to nucleosynthetic effects. An underabundance of carbon is often observed in DLAs (see Sec.\ \ref{Abundances}). Aluminium, on the other hand, might suffer from uncertainties in the shape of the ionising spectrum or be subject to an intrinsic underabundance compared to a solar abundance pattern \citep{Crighton2013,Rolleston2003}. Despite the underabundance, the ratios of Al\,{\sc ii}/ Al\,{\sc iii} and N\,{\sc i}/ N\,{\sc ii} are modelled correctly.

\textit{Model 8.3} -- This model is based on a metallicity of 2.75\% solar and a hydrogen density of $\log n_{\text{H}}= -2.0$. It indicates the low density limit of the range where the Al\,{\sc ii}/ Al\,{\sc iii} and N\,{\sc i}/ N\,{\sc ii} ratios match the observations, whereas the high density limit is indicated by Model 8.2. The O\,{\sc i} column density is well reproduced. Too little S\,{\sc ii} is produced. All other column densities are overproduced compared to their measured values. These underabundances can qualitatively be explained by nucleosynthesis and dust depletion. The ratios of Al\,{\sc ii}/ Al\,{\sc iii} and N\,{\sc i}/ N\,{\sc ii} are modelled correctly.

\textit{Model 8.4} -- This model is based on a metallicity of 6\% solar and a hydrogen density of $\log n_{\text{H}}= -3.0$. The metallicity in this model does not correspond to the measured value of [O\,{\sc i}/H\,{\sc i}], but it produces the column densities of O\,{\sc i} and S\,{\sc ii} correctly. However, it overproduces all of the other column densities. An underabundance can again be qualitatively explained by nucleosynthesis and dust depletion. The N\,{\sc i}/ N\,{\sc ii} ratio is reproduced correctly, but the Al\,{\sc ii}/ Al\,{\sc iii} ratio is not. Besides, a hydrogen density of $\log n_{\text{H}}= -3.0$ seems rather low for a sub-DLA system. Such a very low gas density together with the total gas column density would imply a linear size of this absorption component of several dozen kpc, which is implausible.

\begin{figure}
 \resizebox{\hsize}{!}{\includegraphics{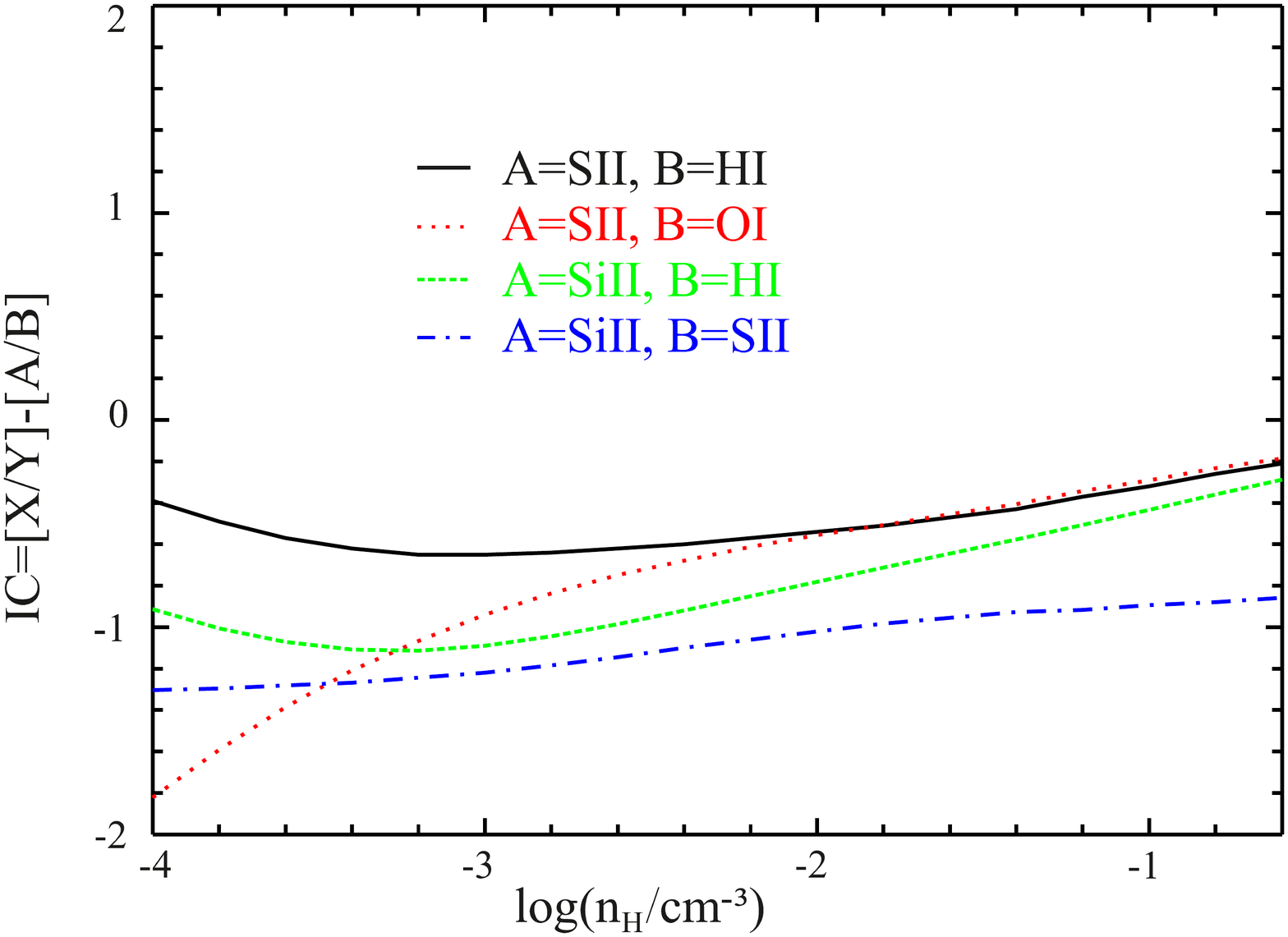}}
\caption {Ionisation correction factors for some elemental column densities in component 8 derived by Cloudy. The model is based on cosmic microwave background radiation and a Haardt \& Madau (2001) UV background, relative solar chemical abundances, and a metallicity of 2.75\% solar. The correction factors are significant, but they cannot explain the full discrepancy between sulphur and the other elements measured.} \label{IC8}
\end{figure}

Fig.\ \ref{IC8} presents the correction factors that can be derived by the Cloudy models. The correction factors show the deviation between the real intrinsic abundances in the gas and the abundances indicated by the dominant ionisation stages:\\

 IC= [X/Y] - [A/B],\\ 

\noindent i.e. they give an estimate of the systematic bias that occurs when ionisation effects are neglected. These corrections are already included in Fig.\ \ref{Cloudy plots}, because Cloudy considers ionisation effects when calculating the column densities. Therefore, the discrepancy between the expected and the measured values cannot be explained by ionisation even though the ionisation corrections for S\,{\sc ii} and Si\,{\sc ii} are significant. In order to estimate the uncertainties caused by the choice of the ionising spectrum, we have  performed more Cloudy calculations for alternative ionising backgrounds. They can be found in the Appendix. None of them can reproduce the observed column density pattern. We note that the significant correction factors found here are in contradiction to previous studies of higher column density DLAs, which find that corrections are generally small or even below measurement errors \citep[e.g.][]{Vladilo2001}.

As none of the models can fully explain the observed abundance pattern, we are forced to conclude that either there is a large intrinsic overabundance of sulphur in this absorption component, or the absorption feature that falls on the S\,{\sc ii} $\lambda 1259$ line at $z=1.839$ is not due to sulphur, but to a blend from another line, or there is unresolved substructure and multi-phase gas in this absorption component, so the Cloudy model cannot correctly reproduce the column-density ratios of the different ions.

Fig.\ \ref{Cloudy plots}c) and d) show similar Cloudy models for component 6. This component is especially interesting, because we can give an exact value for Al\,{\sc iii} and not just an upper limit. The H\,{\sc i} column density in this component is estimated to be $\log N$(H\,{\sc i}) $= 18.7$, based on the (O\,{\sc i}/H\,{\sc i}) ratio.  We find three models that partly match the observations. The deviations from the measured values are presented in Table\ \ref{Deviations component 6}.

\textit{Model 6.1} -- This model is based on 2.75\% solar metallicity and a hydrogen density of $\log n_{\text{H}}= -1.2$. It indicates the high density limit of the range where the O\,{\sc i} column density matches the observations whereas Model 6.2 indicates the low density limit. Model 1 also reproduces the C\,{\sc ii} and Si\,{\sc ii} column density, but it underproduces magnesium and sulphur and overproduces iron. The N\,{\sc i}/ N\,{\sc ii} ratio agrees with the upper limits, but the Al\,{\sc ii}/ Al\,{\sc iii} is not predicted correctly. 

\textit{Model 6.2} -- This model is based on 2.75\% solar metallicity and a hydrogen density of $\log n_{\text{H}}= -1.8$. It correctly reproduces the O\,{\sc i} and Mg\,{\sc ii} column density. It does not produce enough S\,{\sc ii} column density. The rest of the ionic column densities are overproduced. Again, the N\,{\sc i}/ N\,{\sc ii} ratio agrees with the upper limits, but the Al\,{\sc ii}/ Al\,{\sc iii} is not predicted correctly. 

\textit{Model 6.3} -- This model is based on 4.7\% solar metallicity and a hydrogen density of $\log n_{\text{H}}= -2.4$. It correctly reproduces the O\,{\sc i} and S\,{\sc ii} column density. All other ions are overproduced which could be caused by nucleosynthetic or dust depletion effects. The Al\,{\sc ii}/ Al\,{\sc iii} and N\,{\sc i}/ N\,{\sc ii} ratios are reproduced correctly despite the underabundance of aluminium and nitrogen. 

Most interestingly, Fig.\ \ref{Cloudy plots}d) indicates that sulphur is {\it not} overabundant in this sub-component if we assume an influence of nucleosynthesis and dust depletion and a metallicity that does not agree with the measured [O\,{\sc i}/H\,{\sc i}]. 

We note that, in previous studies, Al\,{\sc iii} often shows a different behaviour than the weakly ionised species. \citet{Vladilo2001} have suggested that Al\,{\sc iii} might originate in a region different from the one where the weakly ionised species arise. In that case, the Al\,{\sc ii}/Al\,{\sc iii} ratio would stand for the relative contribution of these separate regions rather than the ionisation state of a single region. On that account, a mismatching Al\,{\sc ii}/Al\,{\sc iii} ratio does not necessarily have to be a criterion for exclusion for a model.

\subsection{S\,{\sc ii} absorption}\label{SII absorption}

\begin{figure}
 \resizebox{\hsize}{!}{\includegraphics{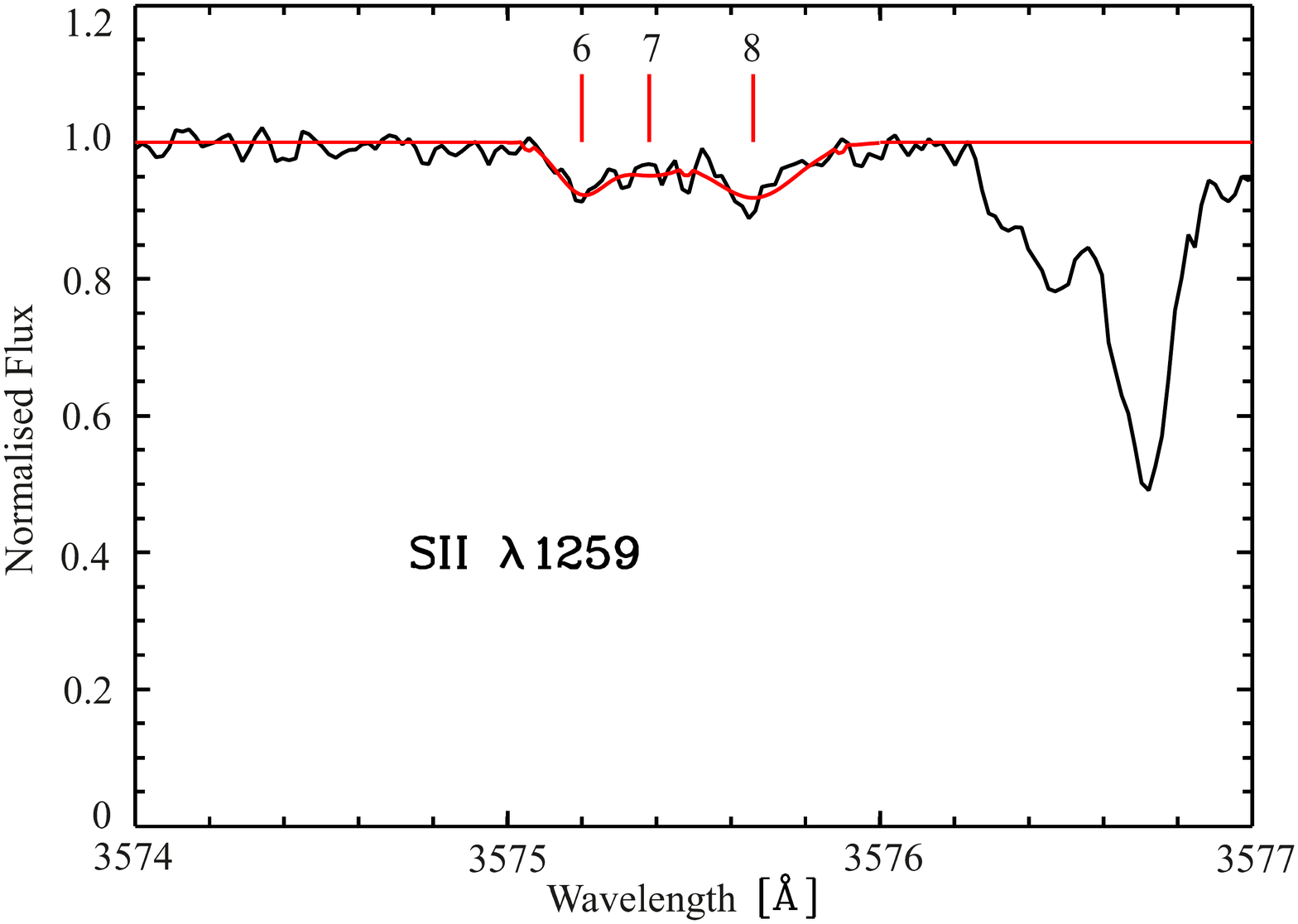}}
\caption {Multi-component Voigt profile fit of the absorption line of S\,{\sc ii} $\lambda$1259 in the sub-DLA at $z=1.839$. The data are black and the fit is red. The velocity components used in the fits are indicated by vertical tick marks. Only the main components have been fitted and they are numbered according to the velocity components of the other weakly ionised species. Parameters of the fit can be found in Table\ \ref{Fit results}.} \label{FitSII}
\end{figure}

Accurate measurements of sulphur in interstellar and intergalactic gas are particularly desirable,
because S is a tracer for the $\alpha$-abundance in the gas and this element is undepleted in dust.
For example, \citet{Nissen2004} list sulphur measurements in DLAs from redshift $\approx1.8 - 3.4$, which they
culled from the literature. \citet{Som2013} measure (among other things) sulphur abundances in sub-DLA systems with redshifts between $\approx1.7 - 2.3$.  \citet{Sparre2013} show measurements of sulphur and other metals in QSO and gamma-ray burst spectra up to $z \approx 6$, also culled from the literature. \citet{Centurion2000} report on sulphur measurements in DLAs at $z\geq2$. On the other side of the redshift scale, \citet{Richter2013} and \citet{Fox2013} have studied S\,{\sc ii} abundances in the Magellanic Stream. 

High metallicity in general is observed sometimes in intervening absorption line systems. \citet{Nissen2004} report on some similarly high sulphur abundances in DLAs. \citet{Noterdaeme2010}, \citet{Peroux2006}, and \citet{Prochaska2006} even find super-solar metallicities in sub-DLAs and Super Lyman Limit Systems, respectively ([S/H]$= 0.15 \pm 0.13$, [Zn/H]$=0.61$, and [S/H]$= 0.79$, respectively). Considering the measured [O\,{\sc i}/H\,{\sc i}], a generally high metallicity does not seem to be the case here. However, in the system presented here, the sulphur abundance appears to be much higher than for the other $\alpha$-elements (see Table\ \ref{Element abundances table}). Interestingly, even though \citet{Peroux2006} do not measure the sulphur abundance in their system, they find evidence of a non-solar abundance pattern as well.  

Another point to consider is the [O\,{\sc i}/Fe\,{\sc ii}] ratio that does not show the expected $\alpha$-enhancement (see Sec.\ \ref{Abundances}). Sulphur, on the other hand, is overabundant compared to iron, but we note that \citet{Nissen2004} also do not find evidence of $\alpha$-enhancement in their collection of DLAs from the literature. This can be interpreted as evidence for low and intermittent rates of star formation over the history
of the observed absorption systems \citep{Calura2003}. In this scenario, the overall metallicity grows slowly with time and
the abundance of iron-peak elements can approach the abundance of $\alpha$-elements in periods without extensive star
formation. Such a trend can be observed locally in dwarf irregular and dwarf spheroidal galaxies, whose stars also do not
show $\alpha$-enhancement at low metallicities \citep[e.g.][]{Shetrone2003,Tolstoy2003}.
 
As we have discussed in Sec.\ \ref{Cloudy}, photoionisation cannot be the origin of the
apparent overabundance of sulphur in the absorber at $z=1.839$, but the Cloudy models suggest 
that the origin for this discrepancy lies in the 8th 
absorption component, where the S\,{\sc ii} $\lambda1259$ absorption is particularly strong.
This is an unexpected trend that clearly needs further exploration.
For this reason, we will discuss in the following different scenarios that may explain
the apparently high sulphur abundance in this system.  

Fig.\ \ref{FitSII} shows a zoom-in of the absorption profile of S\,{\sc ii} $\lambda1259$ together 
with a part of the Si\,{\sc ii} $\lambda1260$ absorption feature. Significant absorption in the $\lambda1259$ line is evident only for the main components 6, 7, and 8. We note that S\,{\sc ii} $\lambda1253$ is unfortunately blended by a Ly $\alpha$ interloper, while possible absorption in
the weak S\,{\sc ii} $\lambda1250$ transition is obviously below the detection limit. 
While the S/N in the spectral region near the S\,{\sc ii} $\lambda1259$ line is mildly lower than in
other regions of the spectrum (see Fig.\, \ref{fitovervelocityscale}), the absorption feature
is clearly detected and cannot be caused by a noise feature.

In principle, there are three possible explanations for the derived overabundance:

\begin{enumerate}

 \item \textit{The detection of sulphur is false positive.} -- In the first place, it is stunning that sulphur 
turns out to be overabundant compared to all other detected elements (see Fig.\ \ref{Element abundances figure}). 
This overabundance cannot be explained by current nucleosynthesis theories, as sulphur is supposedly produced together with the other $\alpha$-elements. As oxygen is also not depleted onto dust grains, sulphur and oxygen should have the same gas-phase abundances. Moreover, the obtained abundance range $-0.86 \le \text{[S\,{\sc ii}/H\,{\sc i}]} \le -0.31$ indicates a metallicity of $\approx14-49$\% solar, which is far above what is expected for protogalactic systems 
at this high redshift and is also inconsistent with the oxygen metallicity. These aspects put the actual detection of sulphur in our absorber into 
question, possibly implying that the absorption feature shown in Fig.\ \ref{FitSII} is not due to S\,{\sc ii}
$\lambda1259$ absorption at $z=1.839$, but is caused by another absorption line at a different redshift. There
is, however, no other prominent absorption system in the spectrum that we are aware of that could be the origin for
this absorption feature near 3575\,\AA . Theoretically, the absorption could also be caused by a very blue component of the Si\,{\sc ii}
$\lambda1260$ line, but this is highly unlikely because no other line indicates clues for any additional absorption components at the considered relative velocity. In addition, the position and relative strength of the absorption components in the absorption feature shown in Fig.\ \ref{FitSII} almost perfectly match the expected component structure of S\,{\sc ii} $\lambda1259$ at $z=1.839$. The probability that a random 
absorption feature from another intervening absorber reproduces by chance the shape of the expected S\,{\sc ii} 
absorption at $z=1.839$ is extremely small. Thus, we are forced to conclude 
that any interpretation of the absorption feature at 3575\,\AA\ other than being S\,{\sc ii} $\lambda1259$ at 
$z=1.839$ is very unlikely. In addition, systematic errors arising from the uncertainties in the atomic data are not expected to play a significant role \citep{Kisielius2014}.

 \item \textit{Sulphur is overabundant in this system due to mixing, non-equilibrium, or nucleosynthetic effects.} --
A sightline passing through intergalactic or galactic gaseous structures only probes a pencil beam, and thus 
it is not guaranteed that the conditions along this sightline are representative for the whole structure. 
Peculiar element abundances could arise from gas underlying dynamic processes 
or gas that is not well mixed. Evidence for this is given by a comparison of the absorption profiles of different elements 
(see Sec.\ \ref{Comparison}). Abundance ratios that cannot be explained by current nucleosynthetic theories have also been 
found by other groups, e.g. by \citet{Fathivavsari2013}, who find sulphur being overabundant compared to oxygen in a sub-DLA, by \citet{Bonifacio2001}, who find sulphur being overabundant compared to oxygen, silicon, and iron even though [S/H]$=-1.04$ 
in the considered DLA, and by \citet{Lehner2008}. On the other hand, super-solar [S/Si] and [S/Fe] ratios could be caused by dust-depletion. Another 
possible explanation for high sulphur abundances could be hypernovae with exploding He-cores, which, according to 
\citet{Nakamura2001}, overproduce sulphur. However, this theory has been disputed by \citet{Nissen2004} who show that 
the general behaviour of sulphur is in good agreement with near-instantaneous production of $\alpha$-elements by 
Type II supernovae. Furthermore, the drawback of the assumption that sulphur is overabundant due to one of the above 
effects is that it provides no explanation for the apparent underabundance of oxygen which leads to a lack of 
$\alpha$-enhancement. By contrast, the sulphur abundance supports the super-solar [Si\,{\sc ii}/Fe\,{\sc ii}] ratio.

 \item \textit{The apparent high sulphur abundance is caused by ionisation effects.} -- The Cloudy models presented 
here are not unique. Different ionisation conditions can result in different abundance ratios. For example, an incident radiation field 
following a power law can produce the observed [S\,{\sc ii}/O\,{\sc i}] ratio. However, this model cannot predict the Al\,{\sc ii}/Al\,{\sc iii} ratio correctly. Evidence for the sulphur abundance being correct might be the observed [$\alpha$/Fe\,{\sc ii}] ratios that only show the expected $\alpha$-enhancement for Si\,{\sc ii} and S\,{\sc ii}, but not for O\,{\sc i}. Furthermore, previous studies of \citet{Richter2005} and \citet{Erni2006} have found an underabundance of carbon in the considered (sub-)DLAs. In the system considered here, we can report an underabundance of carbon in relation to S, Si and Al, but 
carbon is overabundant compared to oxygen and iron. On the other hand, the detected C\,{\sc ii} absorption line is 
likely blended by a weak absorption feature of Al\,{\sc ii} at $z=1.268$. Another argument that supports the sulphur 
abundance found here, is an estimation for the expected sulphur abundance in DLAs taken from 
\citet{Prochaska1999}: They argue that an abundance pattern being consistent with SN Type II enrichment and dust-depletion 
requires [S/Fe]$>0.6$. This is because, on the one hand, sulphur as $\alpha$-element is observed to be overabundant 
relative to Fe in metal-poor halo stars by $\approx 0.3-0.5$ dex and, on the other hand, [Zn/Fe] values resulting from 
dust-depletion suggest [S/Fe]$>0.3$ dex on the basis of depletion alone. Even though \citet{Prochaska1999} find values 
that are inconsistent with this estimation, the expected [S/Fe]$>0.6$ is consistent with our results, but other 
ion ratios (see above) remain inconsistent.

\end{enumerate}

As none of the above three scenarios can explain {\it individually} the unusually high abundance of S\,{\sc ii} in the absorber, we suggest that the observed abundance pattern in this system reflects a particular local environment in the absorbing gas structure where the (combined) effects of a multi-phase nature of the gas \citep[see also][]{Fumagalli2013}, unusual intrinsic abundances, dust depletion, and/or unresolved sub-component structure are relevant. Neutral gas coexisiting with ionised gas can result in peculiar abundance ratios and Cloudy is unable to model multi-phase systems \citep[see][]{Milutinovic2010,Lehner2008,Fox2007b}. Besides, a multi-phase gas structure is motivated by the evidently strong interlacing of the absorption components of highly and weakly ionised species (see Fig.\ \ref{Vergleich}). This point of view is also motivated by previous studies of our Galactic halo, e.g. \citet{Howk2006} who argue that comparing the column densities of the dominant ionisation stages of metals to the column density of neutral hydrogen along a 
line of sight will result in apparently too high metal abundances.

\section{Relevance to other studies} \label{Discussion}

The sub-DLA system at $z=1.839$ is an excellent example for our lack of understanding of the physical and 
chemical conditions in intervening absorption line systems. Our detailed analysis reveals that, despite of
the excellent data quality (or rather because of), the derived abundances for the two $\alpha$-elements
O and S cannot be explained by a standard SN Type II enrichment pattern, even if 
photoionisation effects are included. In the light of the very high resolution used here, discrepancies arise 
that challenge the results of lower-resolution studies on intergalactic absorption line systems. Especially, the summing up of metal 
abundances and the averaging of temperatures and densities over these absorbing structures -- an often used approach for the interpretation of multi-component absorbers -- seems questionable, taking into account the substantial differences that we find between single velocity components of the same absorber. 
In contrast to current theories, the abundance pattern in this system is neither completely in 
line with a SN Type II enrichment pattern, nor with an ISM-like dust-depletion pattern, nor with a combination of 
the two if we take into consideration the puzzling S\,{\sc ii} absorption. 
On the other hand, ignoring the S\,{\sc ii} absorption leaves us with a striking underabundance of oxygen, evident 
from Fig.\ \ref{Element abundances figure}. This would result in a lack of $\alpha$-enhancement, which would be commonly 
expected for intervening absorption systems, and requires strong depletion values to fit the Cloudy models, which seems unlikely.

As has been pointed out in Sec.\ \ref{SII absorption}, peculiar sulphur abundances have also been found by \citet{Bonifacio2001}, and \citet{Fathivavsari2013} have derived abundances of $\alpha$-elements not being in line with each other in sub-DLA systems. Other detailed studies of multi-component absorbers at high redshift provide further evidence of significant discrepancies between a standard SN Type II enrichment pattern and measured gas-phase abundances in {\it individual absorption components}. For instance, \citet{Richter2005} have studied a very complex sub-DLA at $z=2.187$ towards
the quasar HE\,0001$-$2340, finding substantial overabundance of several heavy elements (including phosphorus and aluminum) 
in two absorption components, even  after applying an ionisation correction. These authors conclude that
the gas belonging to these components is possibly enriched {\it locally} by the supernova ejecta from one or more massive 
stellar clusters. The lack of mixing of heavy elements in the interstellar and circumgalactic gas of galactic and protogalactic 
structures at high redshift is not unexpected, however, because the assembly and evolution of galaxies in that epoch is boosting and coming along with intense (but spatially irregular) star-formation episodes on time scales that can be shorter than the mixing-time scales of metals in the gas. In addition to the chemical enrichment, the local ionisation conditions may also vary dramatically in the presence of star-forming regions and may strongly deviate from a standard UV background model as commonly used to model intervening absorbers at high $z$. In the light of these arguments, the peculiar abundances found by us in the sub-DLA at $z=1.839$ towards the quasar B1101$-$26 could be yet another example for a rather {\it local}
phenomenon in a protogalactic structure related to star-formation activity \citep[see e.g.][]{Crighton2013,Prochter2010,D'Odorico2001}.

In general, an inhomogeneous distribution of metals (including dust) and irregular ionisation conditions in sub-DLAs and DLAs 
might be major problems for the interpretation of metal abundances in these absorbers. Metal abundances derived 
from a single sightline may not be representative at all for the host galaxy of the absorber and averaging metal abundances
over several velocity components will lead to meaningless results if the metal distribution is patchy within the absorber.
It is also important to note that such metal-mixing issues are not restricted to high redshifts.
One particular prominent example for an inhomogeneous distribution of heavy elements in a sub-DLA at $z=0$ is the 
Magellanic Stream (MS), a massive tidal gas structure located at $\sim 50-60$ kpc distance from the Milky Way that has been ripped
off the Magellanic Clouds $\sim 1-2$ Gyr \citep{Wannier1972,Gardiner1996}. In a recent study, 
\citet{Richter2013} and \citet{Fox2013} have found that the metallicity in the MS varies by a factor of 5 between the main body of the Stream ($0.1$ solar) and the region that is close to the Magellanic Clouds ($0.5$ solar). 
For the MS sightline towards Fairall\,9 \citet{Richter2013} used sulphur as a proxy for the $\alpha$-abundance in the gas. Interestingly, they determine a sulphur abundance in the MS towards Fairall\,9 that is {\it higher} than the present-day sulphur abundance in the Magellanic Clouds (their Fig.\,9), suggesting a local $\alpha$ (sulphur) enrichment by massive stars before the gas was separated from the Magellanic Clouds and incorporated into the MS.
The high sulphur abundance is an interesting similarity between the high-redshift sub-DLA towards the quasar B1101$-$26 
and a very local tidal gas stream, indicating that our understanding of the metal-abundance pattern in gaseous structures that 
have been enriched locally by massive stars is yet incomplete.

We conclude that the aspect of abundance anomalies in sub-DLAs and DLAs at low and high redshift, such as
presented here, needs further investigation to better understand enrichment and mixing processes 
in gas at high redshift and to constrain the effect of local phenomena for the interpretation of
metal absorption in intervening absorbers.

\section{Summary and conclusions}\label{Summary}

We have performed a detailed analysis of the sub-DLA at $z=1.839$ towards the quasar B1101$-$26 
using VLT/UVES spectral data with very high spectral resolution ($R\sim 75,000$) and a very high 
S/N of $>100$. Our results can be summarised as follows:

\begin{itemize}

 \item We identify 11 absorption subcomponents in neutral, weakly, and highly ionised species. Detected species include 
C\,{\sc ii}, C\,{\sc iv}, N\,{\sc ii}, O\,{\sc i}, Mg\,{\sc i}, Mg\,{\sc ii}, Al\,{\sc ii}, Al\,{\sc iii}, Si\,{\sc ii}, 
Si\,{\sc iii}, Si\,{\sc iv}, Fe\,{\sc ii}, and possibly S\,{\sc ii}. These components span a restframe velocity range of 
$\approx 200$\,km s$^{-1}$. The fit of the Lyman $\alpha$ absorption line yields a column density of neutral hydrogen 
of $\log N(\text{H\,{\sc i}}) = 19.48 \pm 0.01$.

 \item The metallicity of this system, as traced by [O\,{\sc i}/H\,{\sc i}], is $-1.56 \pm 0.01$. However, a peculiar abundance ratio between 
oxygen and sulphur is found with $0.69 \le $ [S\,{\sc ii}/O\,{\sc i}] $\le 1.26$. Concludingly, sulphur indicates a much higher metallicity than oxygen.

 \item For the nitrogen-to-oxygen ratio, we derive an upper limit of [N\,{\sc i}/O\,{\sc i}] $\le -0.65$, which suggests that the absorber is chemically young. 
This conclusion is also supported by a supersolar $\alpha$/Fe ratio of [Si\,{\sc ii}/Fe\,{\sc ii}] $\approx 0.5$. However, this ratio might be affected by dust-depletion.

 \item The abundance pattern in this system is not consistent with SN Type II enrichment combined with the effects of dust-depletion. 
The low oxygen-to-iron ratio ($-0.08 \le \text{[O\,{\sc i}/Fe\,{\sc ii}]} \le -0.05$) implies a lack of $\alpha$-enhancement, the observed $\alpha$-elements 
do not follow each other, and there is no obvious reason for sulphur being overabundant relative to oxygen.

 \item Comparing our study with previous studies on this absorption line system, it is evident that we could resolve the absorption lines of the weakly 
and the highly ionised species into more velocity subcomponents because of the higher resolution of the new UVES data. The detection of 
N\,{\sc ii} and the derivation of safe upper and lower limits for the weak absorption features provide additional information on
the chemical composition of the gas.

 \item We have calculated detailed photoionisation models using Cloudy. They yield a metallicity between 2.7\% and 6.0\% solar and a hydrogen density between $10^{-0.8}$ and $10^{-3.0}$\,cm$^{-3}$. The predicted column densities also suggest a depletion of the refractory 
elements including silicon and an underabundance of nitrogen. Furthermore, the Cloudy models indicate that the bigger part of the hydrogen gas in this system is ionised and only a small fraction is in the neutral gas phase. The Cloudy models cannot, however, explain the unusually high [S\,{\sc ii}/O\,{\sc i}] ratio in this system.

 \item We discuss possible origins for the peculiar overabundance of sulphur in the absorber.
We suggest that the high [S\,{\sc ii}/O\,{\sc i}] ratio is caused by the combination of several relevant effects, such as specific ionisation conditions in multi-phase gas, unusual relative abundances of heavy elements, and dust depletion. The sightline possibly passes a local gas environment that is not well mixed and that might be influenced by star-formation activity in the sub-DLA host galaxy. We discuss the implications of our findings for the interpretation of $\alpha$-element abundances in metal absorbers at high redshift.

\end{itemize}

\begin{acknowledgements}
 The authors would like to thank Michael T. Murphy for providing the reduced VLT/UVES data set of QSO B\,$1101-26$. We also thank an anonymous referee who provided valuable comments that helped to improve the paper. A.F. is grateful for financial support from the Leibniz Graduate School for Quantitative Spectroscopy in Astrophysics, a joint project of the Leibniz Institute for Astrophysics Potsdam (AIP) and the Institute of Physics and Astronomy of the University of Potsdam (UP).
 
\end{acknowledgements}

\bibliographystyle{aa}
\bibliography{bibfile}

\begin{thebibliography}{85}
\expandafter\ifx\csname natexlab\endcsname\relax\def\natexlab#1{#1}\fi

\bibitem[{{Acharova} {et~al.}(2013){Acharova}, {Gibson}, {Mishurov}, \&
  {Kovtyukh}}]{Acharova2013}
{Acharova}, I.~A., {Gibson}, B.~K., {Mishurov}, Y.~N., \& {Kovtyukh}, V.~V.
  2013, \aap, 557, A107

\bibitem[{{Arnett}(1971)}]{Arnett1971}
{Arnett}, W.~D. 1971, \apj, 166, 153

\bibitem[{{Asplund} {et~al.}(2009){Asplund}, {Grevesse}, {Sauval}, \&
  {Scott}}]{Asplund2009}
{Asplund}, M., {Grevesse}, N., {Sauval}, A.~J., \& {Scott}, P. 2009, \araa, 47,
  481

\bibitem[{{Boissier} {et~al.}(2003){Boissier}, {P{\'e}roux}, \&
  {Pettini}}]{Boissier2003}
{Boissier}, S., {P{\'e}roux}, C., \& {Pettini}, M. 2003, \mnras, 338, 131

\bibitem[{{Bonifacio} {et~al.}(2001){Bonifacio}, {Caffau}, {Centuri{\'o}n},
  {Molaro}, \& {Vladilo}}]{Bonifacio2001}
{Bonifacio}, P., {Caffau}, E., {Centuri{\'o}n}, M., {Molaro}, P., \& {Vladilo},
  G. 2001, \mnras, 325, 767

\bibitem[{{Calura} {et~al.}(2003){Calura}, {Matteucci}, \&
  {Vladilo}}]{Calura2003}
{Calura}, F., {Matteucci}, F., \& {Vladilo}, G. 2003, \mnras, 340, 59

\bibitem[{{Centuri{\'o}n} {et~al.}(2000){Centuri{\'o}n}, {Bonifacio}, {Molaro},
  \& {Vladilo}}]{Centurion2000}
{Centuri{\'o}n}, M., {Bonifacio}, P., {Molaro}, P., \& {Vladilo}, G. 2000,
  \apj, 536, 540

\bibitem[{{Crighton} {et~al.}(2013){Crighton}, {Hennawi}, \&
  {Prochaska}}]{Crighton2013}
{Crighton}, N.~H.~M., {Hennawi}, J.~F., \& {Prochaska}, J.~X. 2013, \apjl, 776,
  L18

\bibitem[{{Dessauges-Zavadsky} {et~al.}(2001){Dessauges-Zavadsky}, {D'Odorico},
  {McMahon}, {Molaro}, {Ledoux}, {P{\'e}roux}, \&
  {Storrie-Lombardi}}]{Dessauges2001}
{Dessauges-Zavadsky}, M., {D'Odorico}, S., {McMahon}, R.~G., {et~al.} 2001,
  \aap, 370, 426

\bibitem[{{Dessauges-Zavadsky} {et~al.}(2003){Dessauges-Zavadsky},
  {P{\'e}roux}, {Kim}, {D'Odorico}, \& {McMahon}}]{Dessauges2003}
{Dessauges-Zavadsky}, M., {P{\'e}roux}, C., {Kim}, T.-S., {D'Odorico}, S., \&
  {McMahon}, R.~G. 2003, \mnras, 345, 447

\bibitem[{{D'Odorico} \& {Petitjean}(2001)}]{D'Odorico2001}
{D'Odorico}, V. \& {Petitjean}, P. 2001, \aap, 370, 729

\bibitem[{{Erni} {et~al.}(2006){Erni}, {Richter}, {Ledoux}, \&
  {Petitjean}}]{Erni2006}
{Erni}, P., {Richter}, P., {Ledoux}, C., \& {Petitjean}, P. 2006, \aap, 451, 19

\bibitem[{{Fabbian} {et~al.}(2009){Fabbian}, {Nissen}, {Asplund}, {Pettini}, \&
  {Akerman}}]{Fabbian2009}
{Fabbian}, D., {Nissen}, P.~E., {Asplund}, M., {Pettini}, M., \& {Akerman}, C.
  2009, \aap, 500, 1143

\bibitem[{{Fathivavsari} {et~al.}(2013){Fathivavsari}, {Petitjean}, {Ledoux},
  {Noterdaeme}, {Srianand}, {Rahmani}, \& {Ajabshirizadeh}}]{Fathivavsari2013}
{Fathivavsari}, H., {Petitjean}, P., {Ledoux}, C., {et~al.} 2013, \mnras, 435,
  1727

\bibitem[{{Faucher-Gigu{\`e}re} \& {Kere{\v s}}(2011)}]{Faucher-Giguere2011}
{Faucher-Gigu{\`e}re}, C.-A. \& {Kere{\v s}}, D. 2011, \mnras, 412, L118

\bibitem[{{Fechner} \& {Richter}(2009)}]{Fechner2009}
{Fechner}, C. \& {Richter}, P. 2009, \aap, 496, 31

\bibitem[{{Ferland} {et~al.}(1998){Ferland}, {Korista}, {Verner}, {Ferguson},
  {Kingdon}, \& {Verner}}]{Ferland1998}
{Ferland}, G.~J., {Korista}, K.~T., {Verner}, D.~A., {et~al.} 1998, \pasp, 110,
  761

\bibitem[{{Field} \& {Steigman}(1971)}]{Field1971}
{Field}, G.~B. \& {Steigman}, G. 1971, \apj, 166, 59

\bibitem[{{Fontana} \& {Ballester}(1995)}]{Fontana1995}
{Fontana}, A. \& {Ballester}, P. 1995, The Messenger, 80, 37

\bibitem[{{Fox} {et~al.}(2007{\natexlab{a}}){Fox}, {Ledoux}, {Petitjean}, \&
  {Srianand}}]{Fox2007}
{Fox}, A.~J., {Ledoux}, C., {Petitjean}, P., \& {Srianand}, R.
  2007{\natexlab{a}}, \aap, 473, 791

\bibitem[{{Fox} {et~al.}(2011){Fox}, {Ledoux}, {Petitjean}, {Srianand}, \&
  {Guimar{\~a}es}}]{Fox2011}
{Fox}, A.~J., {Ledoux}, C., {Petitjean}, P., {Srianand}, R., \&
  {Guimar{\~a}es}, R. 2011, \aap, 534, A82

\bibitem[{{Fox} {et~al.}(2007{\natexlab{b}}){Fox}, {Petitjean}, {Ledoux}, \&
  {Srianand}}]{Fox2007b}
{Fox}, A.~J., {Petitjean}, P., {Ledoux}, C., \& {Srianand}, R.
  2007{\natexlab{b}}, \apjl, 668, L15

\bibitem[{{Fox} {et~al.}(2013){Fox}, {Richter}, {Wakker}, {Lehner}, {Howk},
  {Ben Bekhti}, {Bland-Hawthorn}, \& {Lucas}}]{Fox2013}
{Fox}, A.~J., {Richter}, P., {Wakker}, B.~P., {et~al.} 2013, \apj, 772, 110

\bibitem[{{Fritze-v.~Alvensleben} {et~al.}(2001){Fritze-v.~Alvensleben},
  {Lindner}, {M{\"o}ller}, \& {Fricke}}]{Alvensleben2001}
{Fritze-v.~Alvensleben}, U., {Lindner}, U., {M{\"o}ller}, C.~S., \& {Fricke},
  K.~J. 2001, \apss, 276, 1007

\bibitem[{{Fumagalli}(2014)}]{Fumagalli2013}
{Fumagalli}, M. 2014, \memsai, 85, 355

\bibitem[{{Fumagalli} {et~al.}(2011){Fumagalli}, {Prochaska}, {Kasen}, {Dekel},
  {Ceverino}, \& {Primack}}]{Fumagalli2011}
{Fumagalli}, M., {Prochaska}, J.~X., {Kasen}, D., {et~al.} 2011, \mnras, 418,
  1796

\bibitem[{{Gardiner} \& {Noguchi}(1996)}]{Gardiner1996}
{Gardiner}, L.~T. \& {Noguchi}, M. 1996, \mnras, 278, 191

\bibitem[{{Grevesse} \& {Sauval}(1998)}]{Grevesse1998}
{Grevesse}, N. \& {Sauval}, A.~J. 1998, \ssr, 85, 161

\bibitem[{{Henry} {et~al.}(2000){Henry}, {Edmunds}, \&
  {K{\"o}ppen}}]{Henry2000}
{Henry}, R.~B.~C., {Edmunds}, M.~G., \& {K{\"o}ppen}, J. 2000, \apj, 541, 660

\bibitem[{{Howk} {et~al.}(2006){Howk}, {Sembach}, \& {Savage}}]{Howk2006}
{Howk}, J.~C., {Sembach}, K.~R., \& {Savage}, B.~D. 2006, \apj, 637, 333

\bibitem[{{Jenkins}(2009)}]{Jenkins2009}
{Jenkins}, E.~B. 2009, \apj, 700, 1299

\bibitem[{{Kisielius} {et~al.}(2014){Kisielius}, {Kulkarni}, {Ferland},
  {Bogdanovich}, \& {Lykins}}]{Kisielius2014}
{Kisielius}, R., {Kulkarni}, V.~P., {Ferland}, G.~J., {Bogdanovich}, P., \&
  {Lykins}, M.~L. 2014, \apj, 780, 76

\bibitem[{{Lanz} \& {Hubeny}(2003)}]{Lanz2003}
{Lanz}, T. \& {Hubeny}, I. 2003, \apjs, 146, 417

\bibitem[{{Lanzetta} {et~al.}(1995){Lanzetta}, {Wolfe}, \&
  {Turnshek}}]{Lanzetta1995}
{Lanzetta}, K.~M., {Wolfe}, A.~M., \& {Turnshek}, D.~A. 1995, \apj, 440, 435

\bibitem[{{Ledoux} {et~al.}(2003){Ledoux}, {Petitjean}, \&
  {Srianand}}]{Ledoux2003}
{Ledoux}, C., {Petitjean}, P., \& {Srianand}, R. 2003, \mnras, 346, 209

\bibitem[{{Lehner} {et~al.}(2008){Lehner}, {Howk}, {Prochaska}, \&
  {Wolfe}}]{Lehner2008}
{Lehner}, N., {Howk}, J.~C., {Prochaska}, J.~X., \& {Wolfe}, A.~M. 2008,
  \mnras, 390, 2

\bibitem[{{Lu} {et~al.}(1996){Lu}, {Sargent}, {Barlow}, {Churchill}, \&
  {Vogt}}]{Lu1996}
{Lu}, L., {Sargent}, W.~L.~W., {Barlow}, T.~A., {Churchill}, C.~W., \& {Vogt},
  S.~S. 1996, \apjs, 107, 475

\bibitem[{{Matteucci} \& {Greggio}(1986)}]{Matteucci1986}
{Matteucci}, F. \& {Greggio}, L. 1986, \aap, 154, 279

\bibitem[{{McWilliam} {et~al.}(1995){McWilliam}, {Preston}, {Sneden}, \&
  {Searle}}]{McWilliam1995}
{McWilliam}, A., {Preston}, G.~W., {Sneden}, C., \& {Searle}, L. 1995, \aj,
  109, 2757

\bibitem[{{Meiksin}(2009)}]{Meiksin2009}
{Meiksin}, A.~A. 2009, Reviews of Modern Physics, 81, 1405

\bibitem[{{Meiring} {et~al.}(2009){Meiring}, {Lauroesch}, {Kulkarni},
  {P{\'e}roux}, {Khare}, \& {York}}]{Meiring2009}
{Meiring}, J.~D., {Lauroesch}, J.~T., {Kulkarni}, V.~P., {et~al.} 2009, \mnras,
  397, 2037

\bibitem[{{Milutinovic} {et~al.}(2010){Milutinovic}, {Ellison}, {Prochaska}, \&
  {Tumlinson}}]{Milutinovic2010}
{Milutinovic}, N., {Ellison}, S.~L., {Prochaska}, J.~X., \& {Tumlinson}, J.
  2010, \mnras, 408, 2071

\bibitem[{{M{\o}ller} {et~al.}(2013){M{\o}ller}, {Fynbo}, {Ledoux}, \&
  {Nilsson}}]{Moller2013}
{M{\o}ller}, P., {Fynbo}, J.~P.~U., {Ledoux}, C., \& {Nilsson}, K.~K. 2013,
  \mnras, 430, 2680

\bibitem[{{Morton}(2003)}]{Morton2003}
{Morton}, D.~C. 2003, \apjs, 149, 205

\bibitem[{{Nakamura} {et~al.}(2001){Nakamura}, {Umeda}, {Iwamoto}, {Nomoto},
  {Hashimoto}, {Hix}, \& {Thielemann}}]{Nakamura2001}
{Nakamura}, T., {Umeda}, H., {Iwamoto}, K., {et~al.} 2001, \apj, 555, 880

\bibitem[{{Nissen} {et~al.}(2004){Nissen}, {Chen}, {Asplund}, \&
  {Pettini}}]{Nissen2004}
{Nissen}, P.~E., {Chen}, Y.~Q., {Asplund}, M., \& {Pettini}, M. 2004, \aap,
  415, 993

\bibitem[{{Noterdaeme} {et~al.}(2010){Noterdaeme}, {Petitjean}, {Ledoux},
  {L{\'o}pez}, {Srianand}, \& {Vergani}}]{Noterdaeme2010}
{Noterdaeme}, P., {Petitjean}, P., {Ledoux}, C., {et~al.} 2010, \aap, 523, A80

\bibitem[{{Osterbrock} \& Ferland(2006)}]{Osterbrock2006}
{Osterbrock}, D.~E. \& Ferland, G.~J. 2006, {Astrophysics of Gaseous Nebulae
  and Active Galactic Nuclei} (University Science Books)

\bibitem[{{P{\'e}roux} {et~al.}(2003){P{\'e}roux}, {Dessauges-Zavadsky},
  {D'Odorico}, {Kim}, \& {McMahon}}]{Peroux2003}
{P{\'e}roux}, C., {Dessauges-Zavadsky}, M., {D'Odorico}, S., {Kim}, T.-S., \&
  {McMahon}, R.~G. 2003, \mnras, 345, 480

\bibitem[{{P{\'e}roux} {et~al.}(2007){P{\'e}roux}, {Dessauges-Zavadsky},
  {D'Odorico}, {Kim}, \& {McMahon}}]{Peroux2007}
{P{\'e}roux}, C., {Dessauges-Zavadsky}, M., {D'Odorico}, S., {Kim}, T.-S., \&
  {McMahon}, R.~G. 2007, \mnras, 382, 177

\bibitem[{{P{\'e}roux} {et~al.}(2006){P{\'e}roux}, {Kulkarni}, {Meiring},
  {Ferlet}, {Khare}, {Lauroesch}, {Vladilo}, \& {York}}]{Peroux2006}
{P{\'e}roux}, C., {Kulkarni}, V.~P., {Meiring}, J., {et~al.} 2006, \aap, 450,
  53

\bibitem[{{Petitjean} {et~al.}(2000){Petitjean}, {Srianand}, \&
  {Ledoux}}]{Petitjean2000}
{Petitjean}, P., {Srianand}, R., \& {Ledoux}, C. 2000, \aap, 364, L26

\bibitem[{{Pettini} {et~al.}(2008){Pettini}, {Zych}, {Steidel}, \&
  {Chaffee}}]{Pettini2008}
{Pettini}, M., {Zych}, B.~J., {Steidel}, C.~C., \& {Chaffee}, F.~H. 2008,
  \mnras, 385, 2011

\bibitem[{{Prochaska} {et~al.}(2002){Prochaska}, {Henry}, {O'Meara}, {Tytler},
  {Wolfe}, {Kirkman}, {Lubin}, \& {Suzuki}}]{Prochaska2002b}
{Prochaska}, J.~X., {Henry}, R.~B.~C., {O'Meara}, J.~M., {et~al.} 2002, \pasp,
  114, 933

\bibitem[{{Prochaska} {et~al.}(2006){Prochaska}, {O'Meara}, {Herbert-Fort},
  {Burles}, {Prochter}, \& {Bernstein}}]{Prochaska2006}
{Prochaska}, J.~X., {O'Meara}, J.~M., {Herbert-Fort}, S., {et~al.} 2006, \apjl,
  648, L97

\bibitem[{{Prochaska} \& {Wolfe}(1997)}]{Prochaska1997}
{Prochaska}, J.~X. \& {Wolfe}, A.~M. 1997, \apj, 487, 73

\bibitem[{{Prochaska} \& {Wolfe}(1999)}]{Prochaska1999}
{Prochaska}, J.~X. \& {Wolfe}, A.~M. 1999, \apjs, 121, 369

\bibitem[{{Prochaska} \& {Wolfe}(2002)}]{Prochaska2002}
{Prochaska}, J.~X. \& {Wolfe}, A.~M. 2002, \apj, 566, 68

\bibitem[{{Prochter} {et~al.}(2010){Prochter}, {Prochaska}, {O'Meara},
  {Burles}, \& {Bernstein}}]{Prochter2010}
{Prochter}, G.~E., {Prochaska}, J.~X., {O'Meara}, J.~M., {Burles}, S., \&
  {Bernstein}, R.~A. 2010, \apj, 708, 1221

\bibitem[{{Rafelski} {et~al.}(2011){Rafelski}, {Wolfe}, \&
  {Chen}}]{Rafelski2011}
{Rafelski}, M., {Wolfe}, A.~M., \& {Chen}, H.-W. 2011, \apj, 736, 48

\bibitem[{{Rafelski} {et~al.}(2012){Rafelski}, {Wolfe}, {Prochaska},
  {Neeleman}, \& {Mendez}}]{Rafelski2012}
{Rafelski}, M., {Wolfe}, A.~M., {Prochaska}, J.~X., {Neeleman}, M., \&
  {Mendez}, A.~J. 2012, \apj, 755, 89

\bibitem[{{Rao} \& {Turnshek}(2000)}]{Rao2000}
{Rao}, S.~M. \& {Turnshek}, D.~A. 2000, \apjs, 130, 1

\bibitem[{{Richter} {et~al.}(2013){Richter}, {Fox}, {Wakker}, {Lehner}, {Howk},
  {Bland-Hawthorn}, {Ben Bekhti}, \& {Fechner}}]{Richter2013}
{Richter}, P., {Fox}, A.~J., {Wakker}, B.~P., {et~al.} 2013, \apj, 772, 111

\bibitem[{{Richter} {et~al.}(2005){Richter}, {Ledoux}, {Petitjean}, \&
  {Bergeron}}]{Richter2005}
{Richter}, P., {Ledoux}, C., {Petitjean}, P., \& {Bergeron}, J. 2005, \aap,
  440, 819

\bibitem[{{Rolleston} {et~al.}(2003){Rolleston}, {Venn}, {Tolstoy}, \&
  {Dufton}}]{Rolleston2003}
{Rolleston}, W.~R.~J., {Venn}, K., {Tolstoy}, E., \& {Dufton}, P.~L. 2003,
  \aap, 400, 21

\bibitem[{{Rudie} {et~al.}(2012){Rudie}, {Steidel}, {Trainor}, {Rakic},
  {Bogosavljevi{\'c}}, {Pettini}, {Reddy}, {Shapley}, {Erb}, \&
  {Law}}]{Rudie2012}
{Rudie}, G.~C., {Steidel}, C.~C., {Trainor}, R.~F., {et~al.} 2012, \apj, 750,
  67

\bibitem[{{Salucci} \& {Persic}(1999)}]{Salucci1999}
{Salucci}, P. \& {Persic}, M. 1999, \mnras, 309, 923

\bibitem[{{Savage} \& {Sembach}(1996)}]{SavageSembach1996}
{Savage}, B.~D. \& {Sembach}, K.~R. 1996, \araa, 34, 279

\bibitem[{{Shetrone} {et~al.}(2003){Shetrone}, {Venn}, {Tolstoy}, {Primas},
  {Hill}, \& {Kaufer}}]{Shetrone2003}
{Shetrone}, M., {Venn}, K.~A., {Tolstoy}, E., {et~al.} 2003, \aj, 125, 684

\bibitem[{{Sofia} \& {Jenkins}(1998)}]{Sofia1998}
{Sofia}, U.~J. \& {Jenkins}, E.~B. 1998, \apj, 499, 951

\bibitem[{{Som} {et~al.}(2013){Som}, {Kulkarni}, {Meiring}, {York},
  {P{\'e}roux}, {Khare}, \& {Lauroesch}}]{Som2013}
{Som}, D., {Kulkarni}, V.~P., {Meiring}, J., {et~al.} 2013, \mnras, 435, 1469

\bibitem[{{Sparre} {et~al.}(2014){Sparre}, {Hartoog}, {Kr{\"u}hler}, {Fynbo},
  {Watson}, {Wiersema}, {D'Elia}, {Zafar}, {Afonso}, {Covino}, {de Ugarte
  Postigo}, {Flores}, {Goldoni}, {Greiner}, {Hjorth}, {Jakobsson}, {Kaper},
  {Klose}, {Levan}, {Malesani}, {Milvang-Jensen}, {Nardini}, {Piranomonte},
  {Sollerman}, {S{\'a}nchez-Ram{\'{\i}}rez}, {Schulze}, {Tanvir}, {Vergani}, \&
  {Wijers}}]{Sparre2013}
{Sparre}, M., {Hartoog}, O.~E., {Kr{\"u}hler}, T., {et~al.} 2014, \apj, 785,
  150

\bibitem[{{Suda} {et~al.}(2011){Suda}, {Yamada}, {Katsuta}, {Komiya},
  {Ishizuka}, {Aoki}, \& {Fujimoto}}]{Suda2011}
{Suda}, T., {Yamada}, S., {Katsuta}, Y., {et~al.} 2011, \mnras, 412, 843

\bibitem[{{Tolstoy} {et~al.}(2003){Tolstoy}, {Venn}, {Shetrone}, {Primas},
  {Hill}, {Kaufer}, \& {Szeifert}}]{Tolstoy2003}
{Tolstoy}, E., {Venn}, K.~A., {Shetrone}, M., {et~al.} 2003, \aj, 125, 707

\bibitem[{{van de Voort} {et~al.}(2012){van de Voort}, {Schaye}, {Altay}, \&
  {Theuns}}]{vandeVoort2012}
{van de Voort}, F., {Schaye}, J., {Altay}, G., \& {Theuns}, T. 2012, \mnras,
  421, 2809

\bibitem[{{Viegas}(1995)}]{Viegas1995}
{Viegas}, S.~M. 1995, \mnras, 276, 268

\bibitem[{{Vladilo} {et~al.}(2011){Vladilo}, {Abate}, {Yin}, {Cescutti}, \&
  {Matteucci}}]{Vladilo2011}
{Vladilo}, G., {Abate}, C., {Yin}, J., {Cescutti}, G., \& {Matteucci}, F. 2011,
  \aap, 530, A33

\bibitem[{{Vladilo} {et~al.}(2001){Vladilo}, {Centuri{\'o}n}, {Bonifacio}, \&
  {Howk}}]{Vladilo2001}
{Vladilo}, G., {Centuri{\'o}n}, M., {Bonifacio}, P., \& {Howk}, J.~C. 2001,
  \apj, 557, 1007

\bibitem[{{Wannier} \& {Wrixon}(1972)}]{Wannier1972}
{Wannier}, P. \& {Wrixon}, G.~T. 1972, \apjl, 173, L119

\bibitem[{{Welty} {et~al.}(1999){Welty}, {Hobbs}, {Lauroesch}, {Morton},
  {Spitzer}, \& {York}}]{Weltly1999}
{Welty}, D.~E., {Hobbs}, L.~M., {Lauroesch}, J.~T., {et~al.} 1999, \apjs, 124,
  465

\bibitem[{{Wolfe} {et~al.}(2005){Wolfe}, {Gawiser}, \& {Prochaska}}]{Wolfe2005}
{Wolfe}, A.~M., {Gawiser}, E., \& {Prochaska}, J.~X. 2005, \araa, 43, 861

\bibitem[{{Wolfe} \& {Prochaska}(2000)}]{Wolfe2000}
{Wolfe}, A.~M. \& {Prochaska}, J.~X. 2000, \apj, 545, 591

\bibitem[{{Yates} {et~al.}(2013){Yates}, {Henriques}, {Thomas}, {Kauffmann},
  {Johansson}, \& {White}}]{Yates2013}
{Yates}, R.~M., {Henriques}, B., {Thomas}, P.~A., {et~al.} 2013, \mnras, 435,
  3500

\bibitem[{{Yin} {et~al.}(2011){Yin}, {Matteucci}, \& {Vladilo}}]{Yin2011}
{Yin}, J., {Matteucci}, F., \& {Vladilo}, G. 2011, \aap, 531, A136

\bibitem[{{Zwaan} {et~al.}(2005){Zwaan}, {van der Hulst}, {Briggs},
  {Verheijen}, \& {Ryan-Weber}}]{Zwaan2005}
{Zwaan}, M.~A., {van der Hulst}, J.~M., {Briggs}, F.~H., {Verheijen}, M.~A.~W.,
  \& {Ryan-Weber}, E.~V. 2005, \mnras, 364, 1467

\end{thebibliography}

\newpage
\appendix

\section{Modelling of Lyman $\alpha$ absorption}

\begin{figure}[h]
\resizebox{\hsize}{!}{\includegraphics[angle=-90]{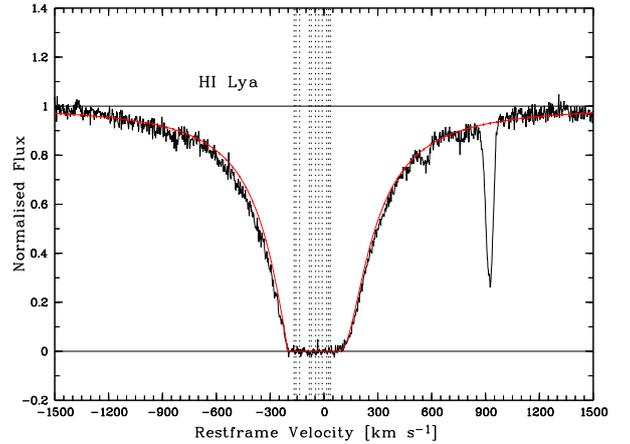}}
\caption{Remodelling of the hydrogen Lyman $\alpha$ absorption line in this absorption system. The absorption line profile is overlayed by the model (red) which is based on the measured O\,{\sc i} column density and the assumption that [O\,{\sc i}/H\,{\sc i}]$=-1.56$ in every component. The vertical, dotted lines indicate the velocity components of the weakly ionised metal species. The shape of the absorption line is reproduced although the outer, weaker components do not give strong constraints on $N$(H\,{\sc i}).\label{ModelHIauskonstOH}}
\end{figure}

\clearpage

\section{Additional Cloudy models} \label{Additional Cloudy plots}

\begin{figure}[htb]
 \noindent
 \mbox{
 \begin{minipage}[t]{0.9\linewidth}
\includegraphics[width=\linewidth]{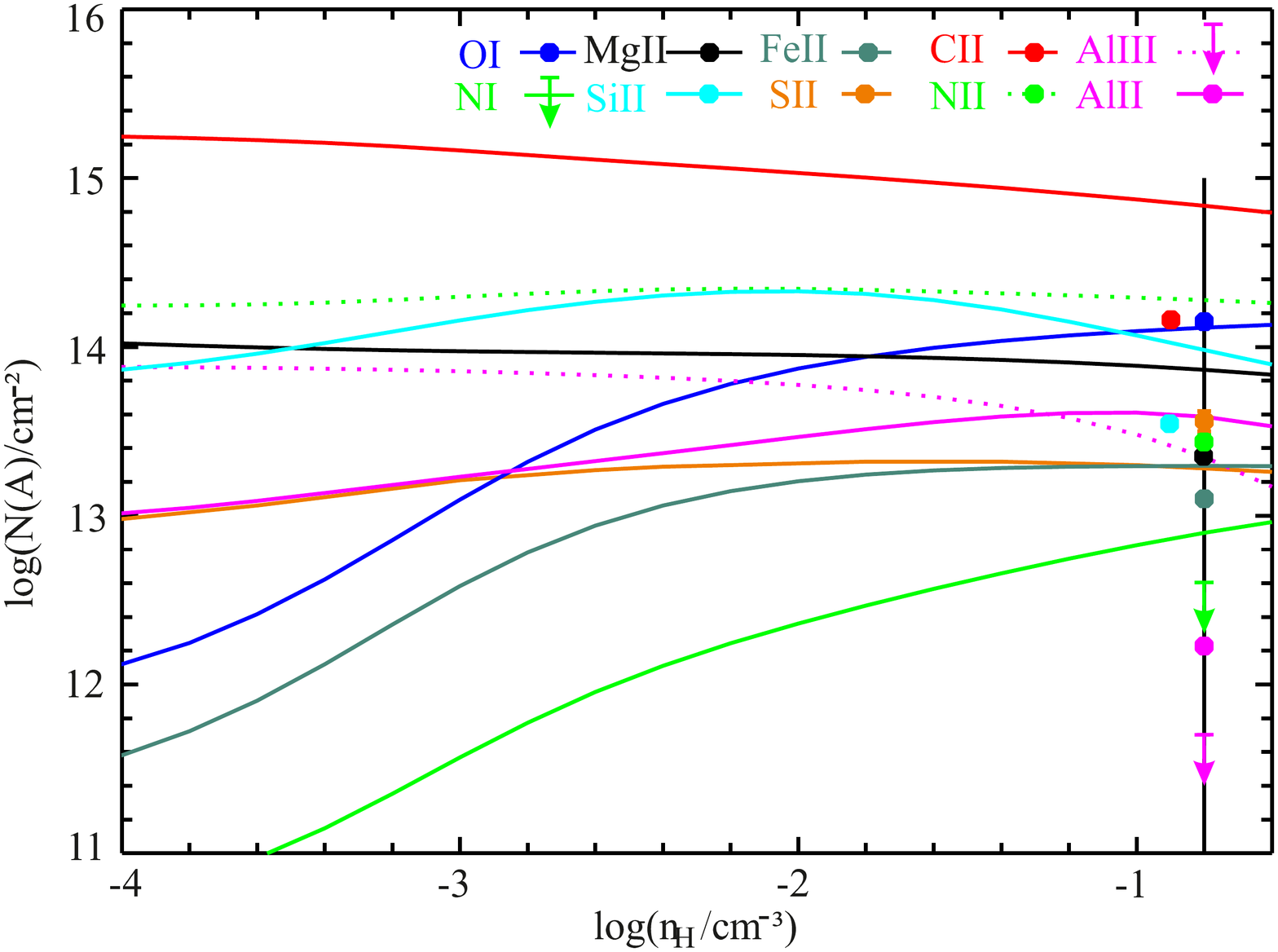}
a) Cloudy Model for component 8 based on the radiation field of an O star. The spectral energy distribution used (Tlusty) has been described in \citet{Lanz2003} and has been prepared by Peter van Hoof for the implementation in Cloudy. The metallicity has been set to 2.75\% solar and the integrated mean intensity of the radiation field at the illuminated face of the cloud has been set to $10^{-2.8}$\,erg\,cm$^{-2}$\,s$^{-1}$.
\end{minipage} \hfill
\begin{minipage}{0.2\linewidth}
 
\end{minipage}
\begin{minipage}[t]{0.9\linewidth}
\includegraphics[width=\linewidth]{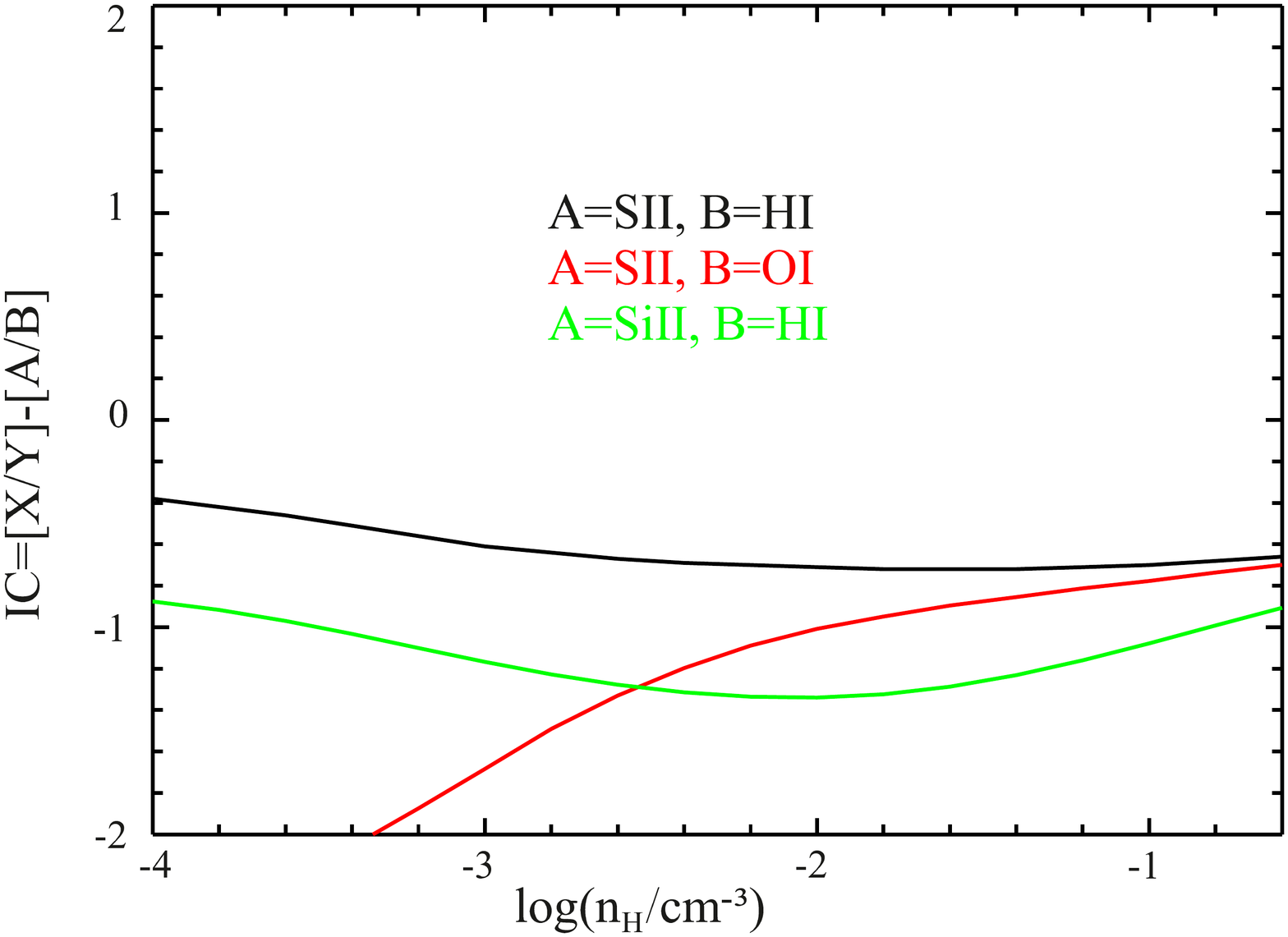}
b) Ionisation correction factors on the basis of a Cloudy model assuming the stellar radiation field of an O star and a metallicity of 2.75\% solar like in a).
\end{minipage}}

\mbox{
\begin{minipage}[t]{0.9\linewidth}
\includegraphics[width=\linewidth]{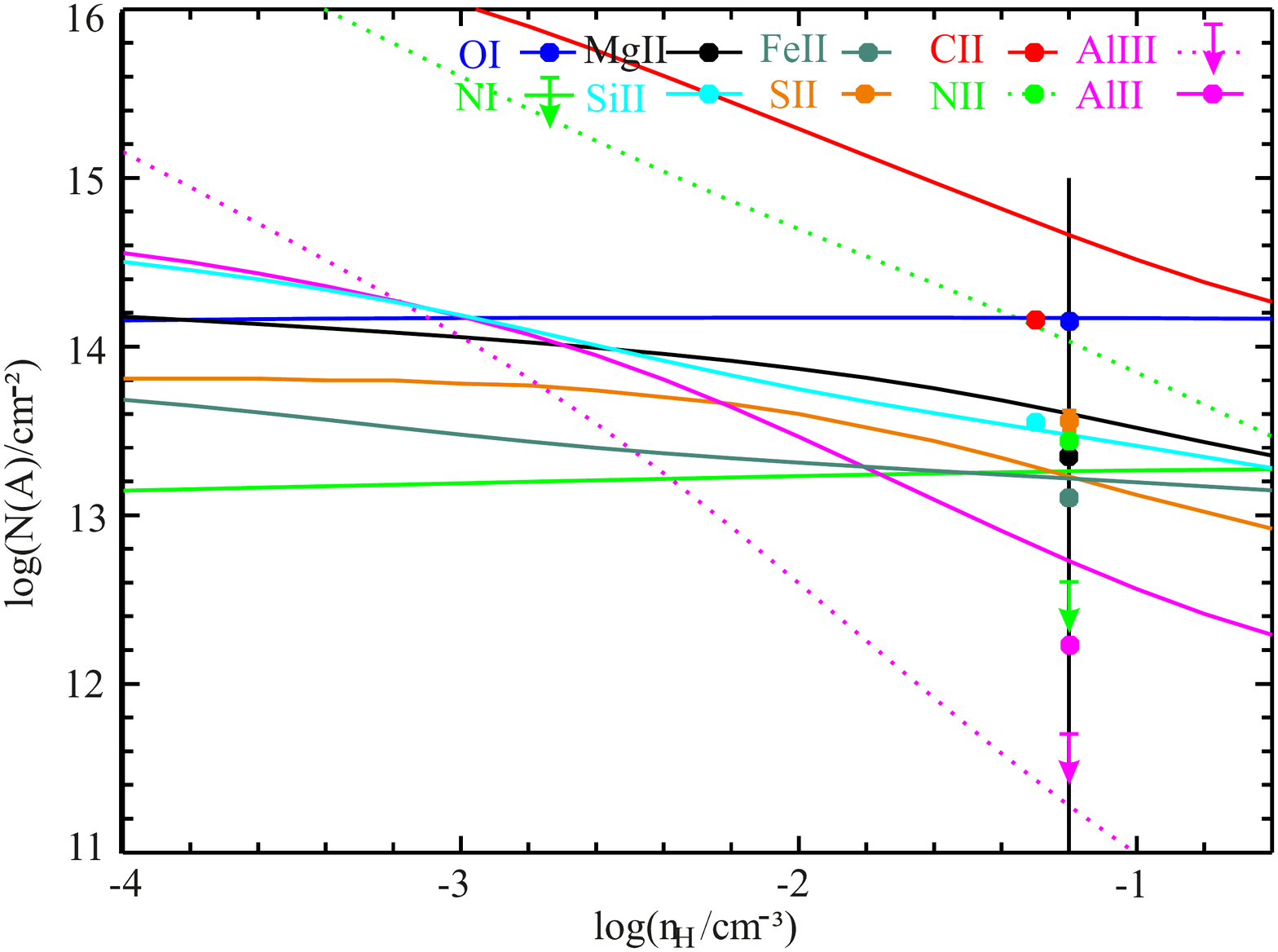}  
c) Cloudy Model for component 8 based on the Galactic interstellar radiation field scaled by the factor 0.1. The metallicity was set to 2.75\% solar.
\end{minipage} \hfill
\begin{minipage}[t]{0.9\linewidth}
 \includegraphics[width=\linewidth]{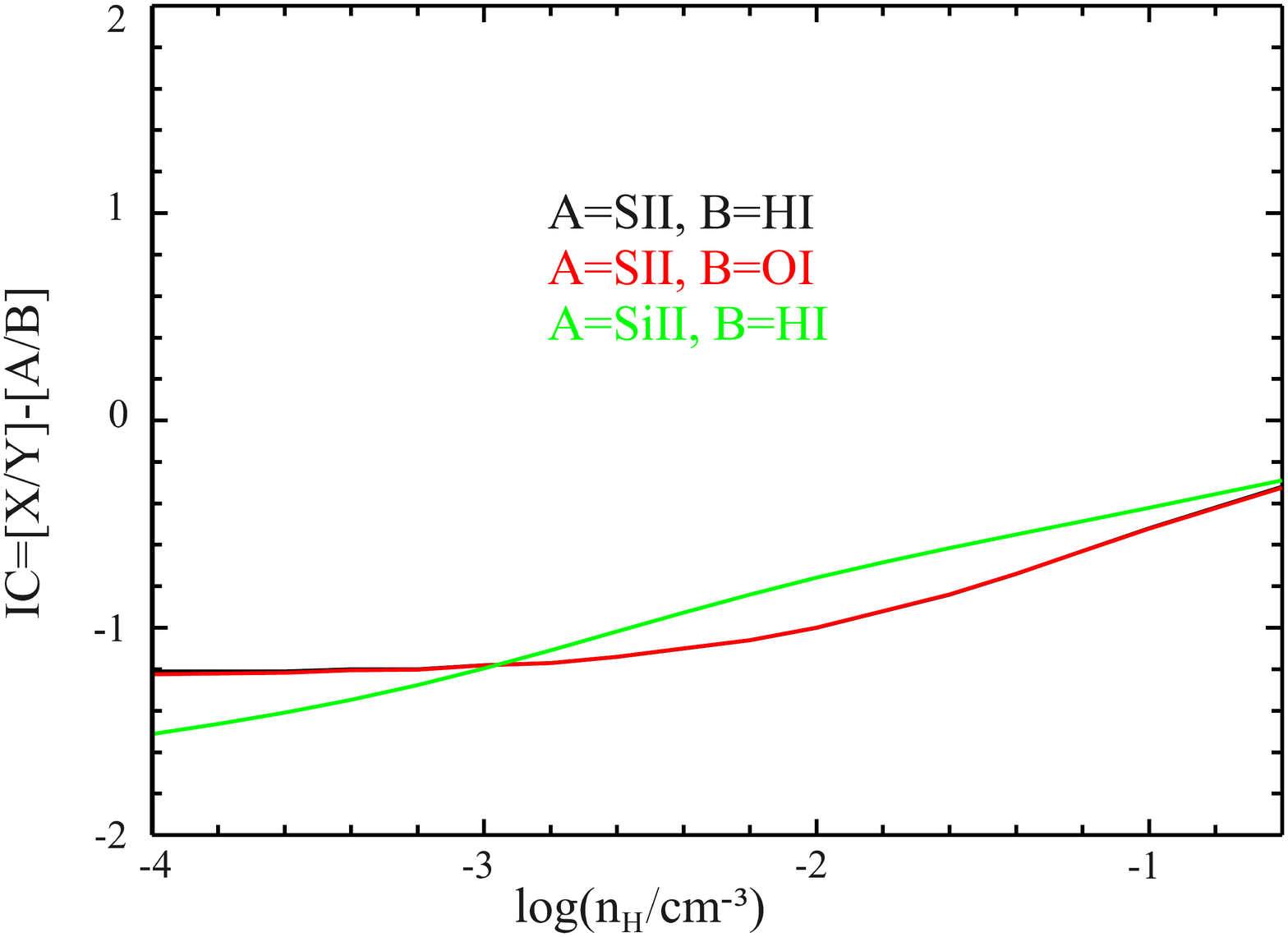} 
\begin{minipage}{0.2\linewidth}
 
\end{minipage}
d) Ionisation correction factors on the basis of a Cloudy model assuming a Galactic interstellar radiation field like in c) and a metallicity of 2.75\% solar.
\end{minipage}}

\caption{\mbox{(c) Cloudy test model for component 8 in the absorption line system at $z=1.839$. 
The lines represent the Cloudy predictions}
\mbox{and the symbols the measured values. The correlation between the symbols 
and the lines is shown in the legends in the upper right} \mbox{corners. Circles correspond to measured values 
with the error bars being smaller than the symbol itself except for sulphur. For Al\,{\sc iii} } \mbox{and N\,{\sc i} upper limits are 
given. The vertical black lines indicate the hydrogen density of our favourite models. The symbols sometimes} \mbox{have an offset for clarity. 
(d) Ionisation correction factors derived by Cloudy based on model c). It is evident that the correction factors} \mbox{are significant, but these (unrealistic) test models, that have been set up to explore the Cloudy parameter range, cannot reproduce} the observed abundance pattern, as expected.
  }
\end{figure}

\end{document}